\documentclass[aps,prl,reprint,amsfonts,amsmath,amssymb,showpacs,groupedaddress,superscriptaddress]{revtex4-2}
\usepackage[colorlinks,urlcolor=blue,linkcolor=blue,anchorcolor=blue,citecolor=blue,bookmarks]{hyperref}
\usepackage{graphicx}
\usepackage{bm}
\usepackage{color}
\usepackage{float}
\usepackage{subfigure}

\begin{document}

\title{Variational Monte Carlo Study of the 1/9 Magnetization Plateau in Kagome Antiferromagnets}
\author{Li-Wei He}
\affiliation{National Laboratory of Solid State Microstructures and Department of Physics, Nanjing University, Nanjing 210093, China}
\author{Shun-Li Yu}
\email{slyu@nju.edu.cn}
\affiliation{National Laboratory of Solid State Microstructures and Department of Physics, Nanjing University, Nanjing 210093, China}
\affiliation{Collaborative Innovation Center of Advanced Microstructures, Nanjing University, Nanjing 210093, China}
\author{Jian-Xin Li}
\email{jxli@nju.edu.cn}
\affiliation{National Laboratory of Solid State Microstructures and Department of Physics, Nanjing University, Nanjing 210093, China}
\affiliation{Collaborative Innovation Center of Advanced Microstructures, Nanjing University, Nanjing 210093, China}

\date{\today}

\begin{abstract}
Motivated by very recent experimental observations of the 1/9 magnetization plateaus in YCu$_3$(OH)$_{6+x}$Br$_{3-x}$ and YCu$_3$(OD)$_{6+x}$Br$_{3-x}$, our study delves into the magnetic field-induced phase transitions in the nearest-neighbor antiferromagnetic Heisenberg model on the kagome lattice using the variational Monte Carlo technique. We uncover a phase transition from a zero-field Dirac spin liquid to a field-induced magnetically disordered phase that exhibits the 1/9 magnetization plateau.  Through a comprehensive analysis encompassing the magnetization distribution, spin correlations, chiral order parameter, topological entanglement entropy, ground-state degeneracy, Chern number and excitation spectrum, we pinpoint the phase associated with this magnetization plateau as a chiral $\mathbb{Z}_3$ topological quantum spin liquid and elucidate its diverse physical properties.
\end{abstract}

\maketitle

The kagome lattice is an exceptional platform for exploring novel many-body states~\cite{PhysRevB.47.5459,PhysRevB.63.014413,PhysRevLett.97.147202,PhysRevB.79.214502,PhysRevB.81.235115,PhysRevLett.106.236802,PhysRevB.83.165118,
PhysRevB.85.144402,nature.562.91,Nature.555.638,science.365.1282,PhysRevLett.125.247002,nature.599.222,
nature.604.59,nature.611.682,nature.612.647,PhysRevLett.98.117205,PhysRevLett.101.117203,Science.332.1173,PhysRevLett.109.067201,PhysRevB.87.060405,PhysRevB.91.020402,PhysRevLett.118.137202,
PhysRevX.7.031020,PhysRevB.95.235107,Sci.Bull.63.1545,Sci.Adv.4.5535,PhysRevB.100.155142,PhysRevB.102.195106,PhysRevB.101.140403,PhysRevB.104.L220408,PhysRevB.76.180407,
PhysRevLett.104.187203,PhysRevLett.98.107204,Nature.492.7429,science.350.655,Nat.Phys.16.469,CPL.34.077502,PhysRevB.98.155127,CPL.37.107503,npj.QM.5.74,Nat.Commun.12.3048,
PhysRevB.105.L121109,arXiv.2311.13089,PhysRevLett.88.057204,PhysRevB.70.100403,JPSJ.79.053707,PhysRevB.83.100405,nat.com.4.2287,PhysRevB.88.144416,PhysRevB.93.060407,
PhysRevB.98.014415,PhysRevB.98.094423,nat.com.10.1229,PhysRevB.107.L220401}, owing to its distinctive lattice and electron structures. In particular, the spin-$1/2$ kagome antiferromagnet with only the nearest-neighbor Heisenberg exhange interactions has attracted significant interest as a promising candidate for realizing the quantum spin liquid (QSL). Although many theoretical studies have suggested that the ground state of this system is likely a QSL~\cite{PhysRevLett.98.117205,PhysRevLett.101.117203,Science.332.1173,PhysRevLett.109.067201,PhysRevB.87.060405, PhysRevB.91.020402, PhysRevLett.118.137202,PhysRevX.7.031020,PhysRevB.95.235107,Sci.Bull.63.1545,Sci.Adv.4.5535,PhysRevB.100.155142,PhysRevB.102.195106,PhysRevB.101.140403,PhysRevB.104.L220408}, there remains a lack of consensus regarding the precise nature of this QSL state.  On the experimental front, the kagome antiferromagnets like herbertsmithite~\cite{PhysRevLett.98.107204,Nature.492.7429,science.350.655,Nat.Phys.16.469}, Zn-barlowite~\cite{CPL.34.077502,PhysRevB.98.155127,CPL.37.107503,npj.QM.5.74,Nat.Commun.12.3048} and YCu$_3$(OH)$_{6+x}$Br$_{3-x}$~\cite{PhysRevB.105.L121109,arXiv.2311.13089} have shown great promise as QSL materials. Moreover,  when subjected to an external magnetic field, the spin-$1/2$ kagome antiferromagnet can also manifest novel quantum states~\cite{PhysRevLett.88.057204,PhysRevB.70.100403,JPSJ.79.053707,PhysRevB.83.100405,nat.com.4.2287,PhysRevB.88.144416,PhysRevB.93.060407,
PhysRevB.98.014415,PhysRevB.98.094423,nat.com.10.1229,PhysRevB.107.L220401}, further highlighting its potential as an ideal platform for exploring exotic quantum states of matter.

Very recently, it was reported experimentally that 1/9 magnetization plateaus were observed in YCu$_3$(OH)$_{6+x}$Br$_{3-x}$ and YCu$_3$(OD)$_{6+x}$Br$_{3-x}$~\cite{arxiv.2310.07989,arxiv.2310.10069,arxiv.2311.11619}. In contrast to the commonly observed $1/3$ magnetization plateaus characterized by classical spin orders in other frustrated antiferromagnets with triangular and honeycomb lattices~\cite{PhysRevB.67.104431,PhysRevB.76.060406,PhysRevLett.108.057205,PhysRevLett.109.267206,PhysRevLett.109.257205,PhysRevLett.110.267201,
nat.com.9.2666,JPSJ.65.2374,nat.phys.19.1883}, this 1/9 plateau phase is a magnetically disordered state. This suggests that the mechanism underlying this phase is fundamentally distinct from the order-by-disorder mechanism that typically gives rise to the $1/3$ magnetization plateaus. However, experimental consensus on certain fundamental characteristics of this phase, such as its gapped or gapless nature, still remains elusive. Theoretically, despite several numerical studies on the spin-$1/2$ kagome antiferromagnetic Heisenberg model have corroborated the existence of the 1/9 magnetization plateau~\cite{nat.com.4.2287,PhysRevB.93.060407,PhysRevB.107.L220401}, its precise nature remains a subject of debate. The density matrix renormalization group (DMRG) calculation has proposed that this plateau phase may correspond to a $\mathbb{Z}_{3}$ spin liquid~\cite{nat.com.4.2287}, whereas the tensor network methods have provided evidence supporting a VBS interpretation~\cite{PhysRevB.93.060407,PhysRevB.107.L220401}. Given these divergent perspectives,  more comprehensive investigations employing a variety of methodologies are necessary to unravel this exotic magnetic phenomenon.

In this letter, we investigate the effect of an external magnetic field on the kagome antiferromagnetic Heisenberg model using the variational Monte Carlo (VMC) method. Without a field, our results reveal that the ground state is a Dirac spin liquid (DSL). This DSL is robust against  weak fields, however, as the field increases beyond a threshold, a new disordered state emerges. This state has a non-zero chiral order parameter and triples the primitive cell. Its magnetization directly jumps onto $M/M_s = 1/9$ ($M_s $ the saturation magnetization) and remains constant over a wide range of field. In this 1/9 plateau phase, the magnetization distribution is uniform. The topological entanglement entropy (TEE) is approximately $\gamma = 1.05$, which is very close to $\ln 3 \approx 1.1$ within numerical error. Based on the relation $\gamma = \ln D \approx \ln 3$, where $D$ is the total quantum dimension, we refer to this exotic state as a chiral $\mathbb{Z}_3$ topological QSL. Moreover, the ground-state degeneracy (GSD) is $9$, implying that this state has an Abelian topological order ($D^2 = 9$). We further unveil the characteristic spin excitation spectrum of this $\mathbb{Z}_3$ QSL, providing key signatures to  identify it directly in experiments.

The kagome antiferromagnetic Heisenberg model in an external magnetic field is written as
\begin{equation}
    H = J\sum_{\langle ij \rangle} \boldsymbol{S}_i \cdot \boldsymbol{S}_j - B_z \sum_i S_i^z,
    \label{eq:model}
\end{equation}
where $\langle ij \rangle$ signifies the sum over nearest-neighbor bonds, $\boldsymbol{S}_i$ represents the spin-$1/2$ operator at site $i$ ($S_i^z$ its $z$-component), $J$ is the exchange interaction, and $B_{z}$ the magnitude of the magnetic field.

\begin{figure}
    \centering
    \subfigure{
    \begin{minipage}[h]{0.4\linewidth}
    \centering
    \includegraphics[width = 1.0\linewidth]{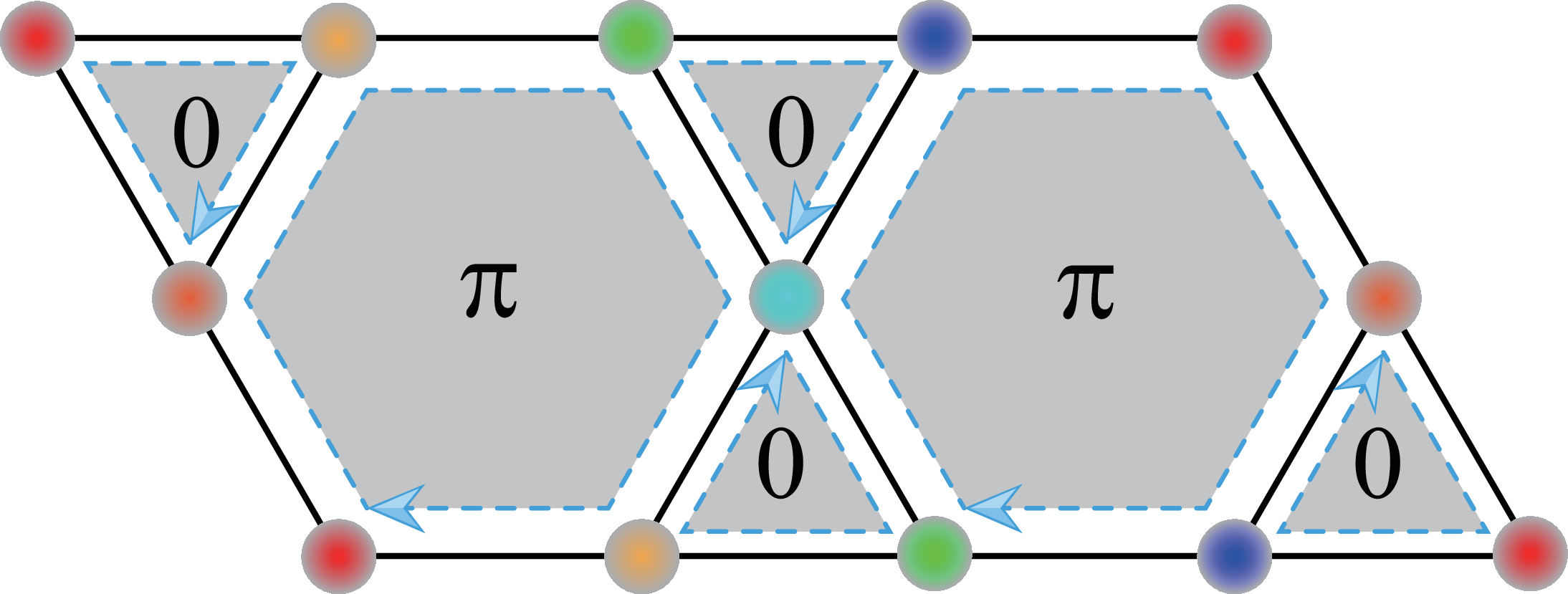}
    \put(-98,2){(a)}
    \label{fig:dsl_ansatz}
    \end{minipage}
    }%
    \subfigure{
    \begin{minipage}{0.55\linewidth}
    \centering
    \includegraphics[width = \linewidth]{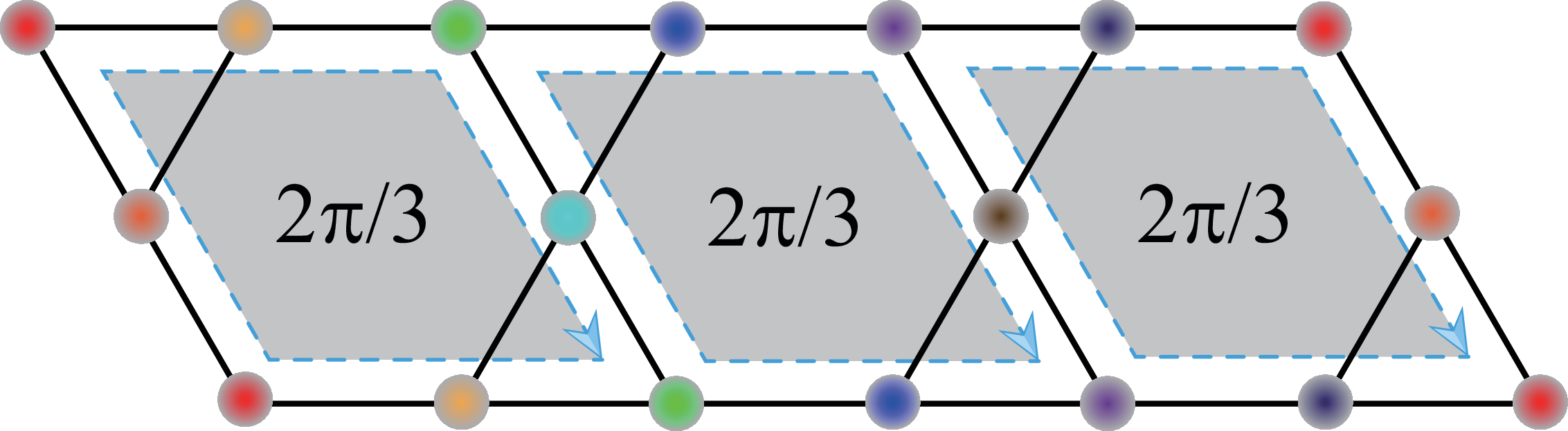}
    \put(-134,2){(b)}
    \label{fig:z3_ansatz}
    \end{minipage}
    }
    \subfigure{
    \begin{minipage}{0.47\linewidth}
    \centering
    \includegraphics[width = \linewidth]{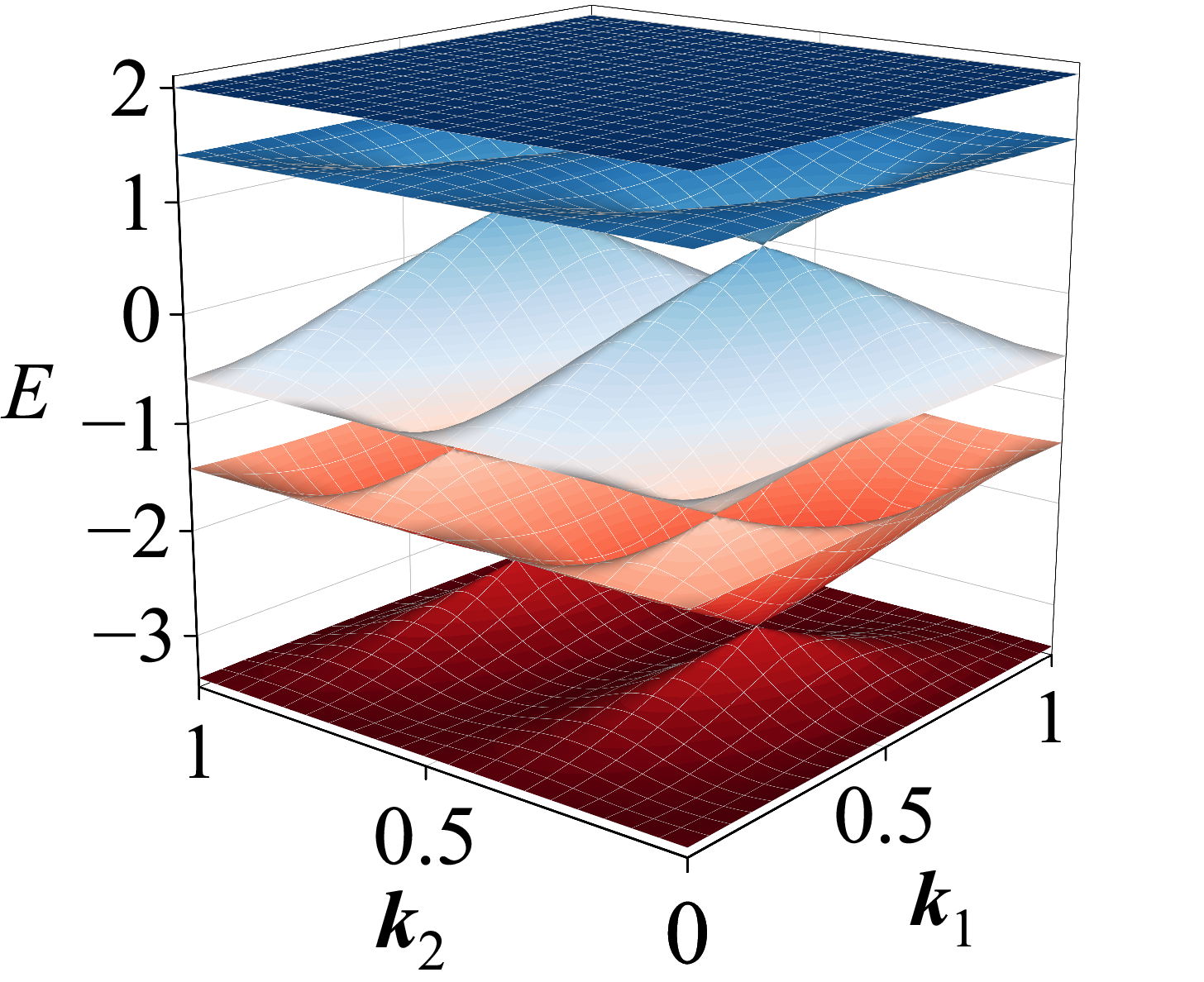}
    \put(-118,90){(c)}
    \label{fig:ek_dsl}
    \end{minipage}
    }%
    \subfigure{
    \begin{minipage}{0.47\linewidth}
    \centering
    \includegraphics[width = \linewidth]{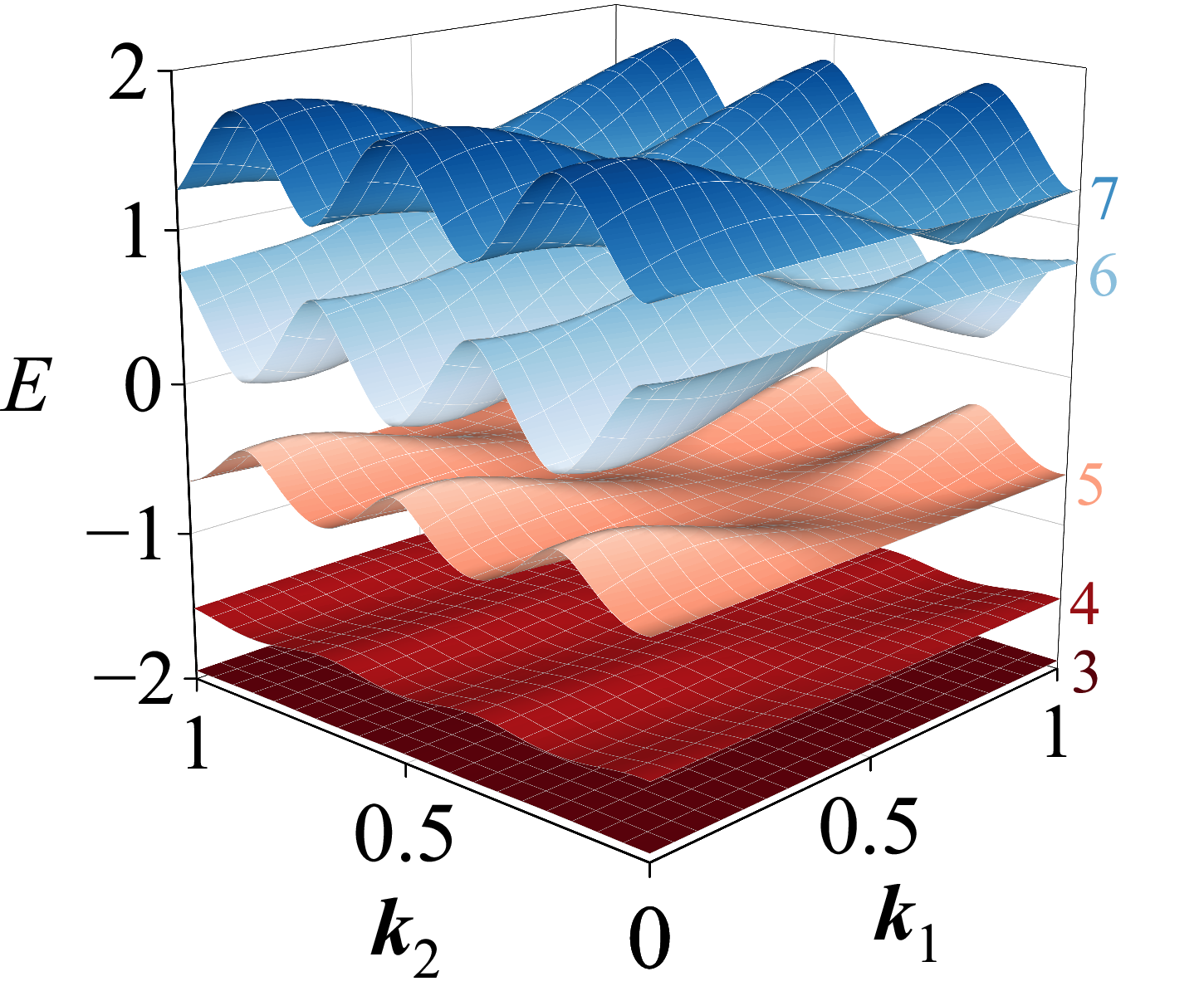}
    \put(-118,90){(d)}
    \label{fig:ek_z3}
    \end{minipage}
    }
    \caption{Mean-field ans\"atzes and spinon band structures. The mean-field ans\"atze for the DSL (a) and $\mathbb{Z}_3$ QSL (b), with flux pattern represented by shaded areas, arrows, and corresponding values.  (c) Complete mean-field spinon band structure for the DSL. (d) Selected mean-field spinon bands (bands 3 to 7) for the $\mathbb{Z}_3$ QSL, using optimal variational parameters derived from VMC calculations at $B_z/J = 0.5$.}
    \label{fig:ansatz_and_ek}
\end{figure}

Following the standard VMC framework, we introduce a fermionic representation for spin operators~\cite{PhysRevB.65.165113,PhysRevB.83.224413,PhysRevB.93.094437,PhysRevLett.127.127205}, {\it i.e.} $\boldsymbol{S}_i = \frac{1}{2} \psi_i^\dagger \boldsymbol{\sigma} \psi_i$ with $\psi_i = (c_{i,\uparrow}, c_{i,\downarrow})^T$, which adhere to the local constraint $\psi_i^\dagger \psi_i= 1$. We then decouple the model (\ref{eq:model}) into a quadratic mean-field Hamiltonian~\cite{supple}:
\begin{equation}
     H_\mathrm{mf} = \sum_{\langle ij \rangle} (t_{ij} \psi_i^\dagger \psi_j + \text{H.c}) - \sum_i \mu \psi_i^\dagger \sigma^z \psi_i,
    \label{eq:full_mf_h}
\end{equation}
where $t_{ij}$ represents the spinon hopping and $\mu$ is the chemical potential that is tuned by the magnetic field. Our analysis of various possible states with spinon pairings reveals that such states are not energetically favorable~\cite{supple}, so the mean-field Hamiltonian considered does not include spinon pairing terms. We construct the variational wave function as $|\Psi(p) \rangle = P_G |GS\rangle_\mathrm{mf}$ with $p = (t_{ij}, \mu)$ embodying the variational parameters, $P_G$ representing the Gutzwiller projector that imposes the strict single-occupation constraint, and $|GS\rangle_\mathrm{mf}$ the ground state of $H_\mathrm{mf}$. The optimization of $p$ is achieved through the minimization of the energy $E(p) = \langle \Psi(p) |H| \Psi(p) \rangle / \langle \Psi(p) | \Psi(p) \rangle$, utilizing the stochastic reconfiguration method~\cite{PhysRevB.61.2599,PhysRevB.64.024512}. For our main results, we utilize lattice sizes of $N = 16 \times 12 \times 3$ for the DSL and $N = 12 \times 12 \times 3$  for the $\mathbb{Z}_3$ QSL and VBS states~\cite{supple}, respectively, unless specified otherwise.

\begin{figure}
    \centering
    \includegraphics[width=1.0\linewidth]{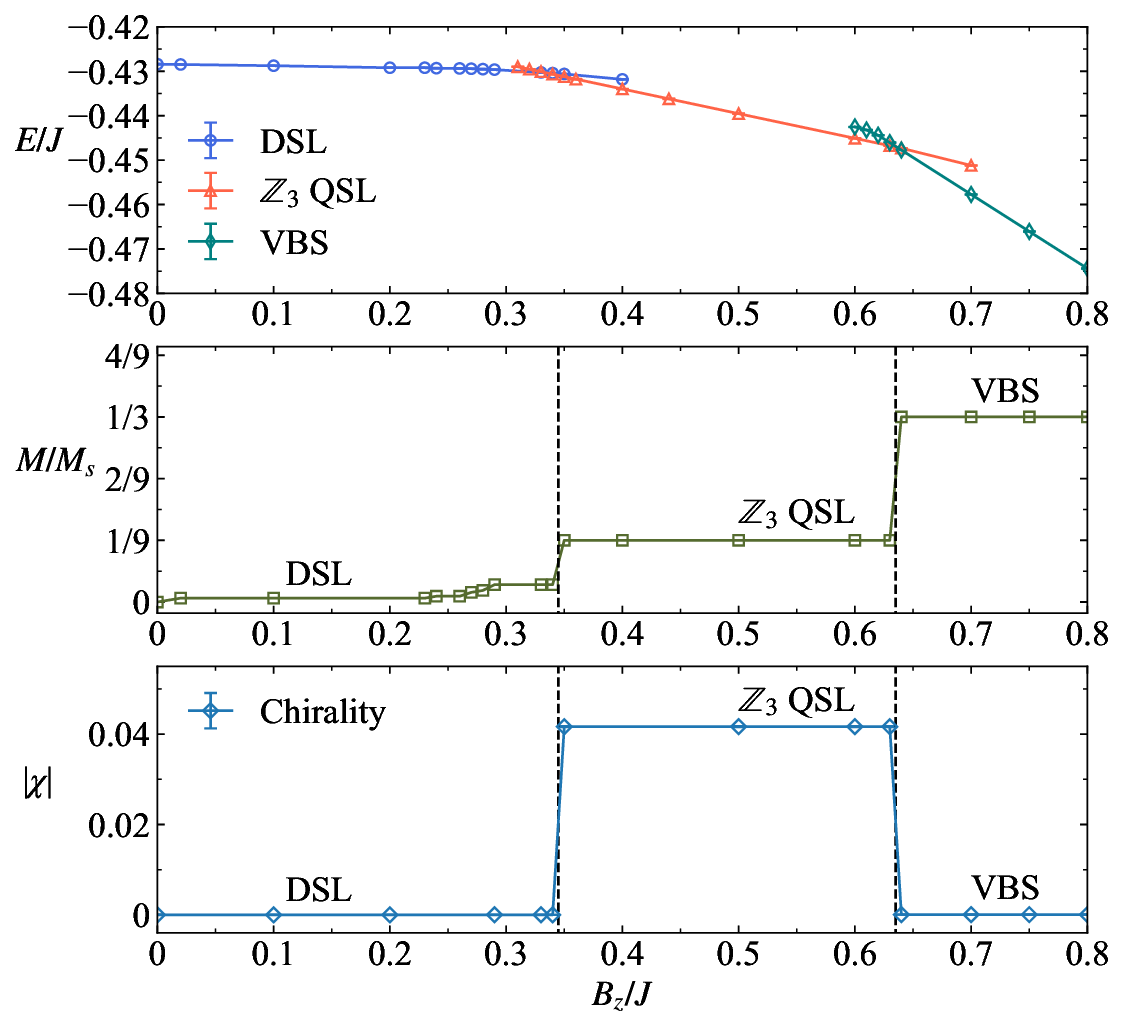}
    \put(-25, 188){(a)}
    \put(-25, 118){(b)}
    \put(-25, 48){(c)}
    \caption{Variational energy $E$ (a), average magnetization $M/M_s$ (b) and chiral order parameter $|\chi|$  (c) as functions of $B_z$.}
    \label{fig:en_m_chi_vs_bz}
\end{figure}

In the regime of low magnetic fields, our VMC calculations reveal that the ground state is a gapless DSL. This state is characterized within the framework of mean-field Hamiltonian (\ref{eq:full_mf_h}) by uniform spinon hopping amplitudes, accompanied by an alternating flux pattern of $0$ and $\pi$ through the triangular and hexagonal plaquettes, as depicted in Fig.~\ref{fig:dsl_ansatz}. As illustrated in Fig.~\ref{fig:ek_dsl}, the mean-field spinon dispersion exhibits characteristic Dirac cones. This DSL is consistent with the results obtained in previous theoretical researches~\cite{PhysRevLett.98.117205,PhysRevB.87.060405,PhysRevLett.118.137202,PhysRevX.7.031020,Sci.Adv.4.5535}. Owing to the vanishing zero-energy density of states for the spinons, the magnetic field must reach a threshold to produce a noticeable magnetization response, as shown in Fig.~\ref{fig:en_m_chi_vs_bz}(b). The false plateau with $M/M_{s}<0.01$ in Fig.~\ref{fig:en_m_chi_vs_bz}(b) is caused by the finite-size effect, it asymptotically approaches to zero with increasing system size~\cite{supple}.

\begin{figure}
    \centering
    \subfigure{
    \begin{minipage}{0.98\linewidth}
    \centering
    \includegraphics[width = 1.0\linewidth]{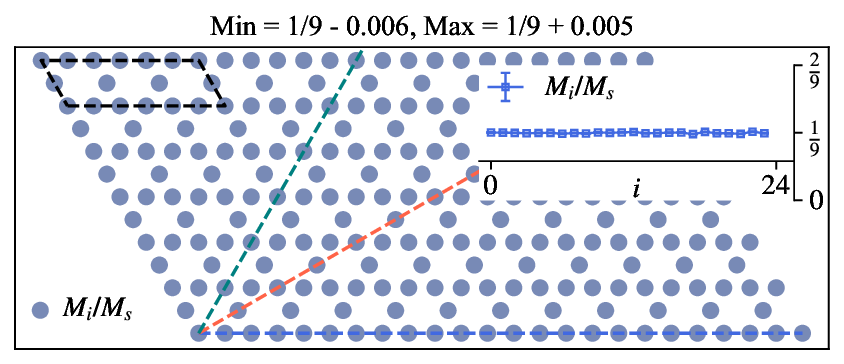}
    \put(-239,95){(a)}
    \label{fig:m_persite}
    \end{minipage}
    }
    \subfigure{
    \begin{minipage}{\linewidth}
    \centering
    \includegraphics[width = 0.97\linewidth]{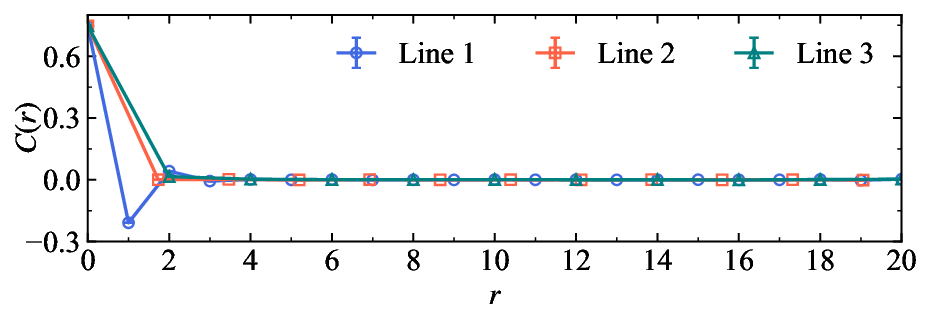}
    \put(-240,75){(b)}
    \label{fig:spin_cor}
    \end{minipage}
    }

    \caption{(a) Distribution of magnetization $M_i/M_s$ in the 1/9 magnetization plateau phase. The black dashed lines highlights the unit cell for this phase. The sphere diameters correspond to the magnetization magnitude at each site. Since the magnetization distribution is almost uniform,  we explicitly provide the minimum and maximum values of magnetization. The magnetization values along a representative direction (indicated by the dashed blue line) are also shown in the inset. (b) Spin-spin correlation functions $C(r)$ along three representative directions, as indicated by the dashed lines with corresponding colors in (a). Here, $C(r) = \sum_{\gamma \in \{x,y,z\}} \langle\Psi| \tilde{S}_{\boldsymbol{r}_0}^\gamma \tilde{S}_{\boldsymbol{r}_0 + r\boldsymbol{\delta}_i}^\gamma|\Psi \rangle$ with $\tilde{S}_{\boldsymbol{r}}^\gamma = S_{\boldsymbol{r}}^\gamma - \langle\Psi| S_{\boldsymbol{r}}^\gamma |\Psi\rangle$, $\boldsymbol{\delta}_i$  being the unit vectors of the three directions and $r$ the distance.}
    \label{fig:mz_and_spin_cor}
\end{figure}

When the field  $B_z$ exceeds $0.35J$, the DSL is no longer energetically favorable, as shown in Fig.~\ref{fig:en_m_chi_vs_bz}(a). The identified phase transition point is in quantitative agreement with that derived from the previous DMRG and tensor network methods~\cite{nat.com.4.2287,PhysRevB.93.060407,PhysRevB.107.L220401}. However, there was a significant divergence in previous studies regarding the nature of the phase after the phase transition, as discussed above. Our VMC calculations reveal that the ground state for $0.35 \alt B_z/J \alt 0.63$ is a $\mathbb{Z}_3$ QSL, whose nature will be discussed in detail subsequently. In the process to search for the most energetically favored state, we have carried out a comprehensive examination of a variety of gauge-inequivalent ans\"atzes, including the uniform resonating-valence-bond state, DSL, chiral spin liquid, $\mathbb{Z}_2$ QSLs with different spinon pairings, $\mathbb{Z}_3$ QSL, and several VBS states~\cite{supple}. For the $\mathbb{Z}_3$ QSL and VBS states, we  extended the unit cell to encompass nine sites, equivalent to three primitive cells of the kagome lattice. For the $\mathbb{Z}_3$ QSL, the amplitudes of the parameter $t_{ij}$ in the mean-field Hamiltonian (\ref{eq:full_mf_h}) and the fluxes within each primitive cell are uniform, making the $3\times1$ and $\sqrt{3}\times\sqrt{3}$ extended unit cells equivalent. Our calculations utilize the $3\times1$ extended unit cell, which facilitates setting the flux in each primitive cell to be $2\pi / 3$, as depicted in Fig.~\ref{fig:z3_ansatz}. To optimize the phases of the parameter $t_{ij}=te^{i\theta_{ij}}$, we incorporate 15 independent $\theta_{ij}$ into our set of variational parameters for the sake of generality. Given that the VBS states can break the translational and rotational symmetries, both the $3\times1$ and $\sqrt{3}\times\sqrt{3}$ extended unit cells are used for the VBS states.

\begin{figure}
    \centering
    \subfigure{
    \begin{minipage}{0.45\linewidth}
    \centering
    \includegraphics[width = \linewidth]{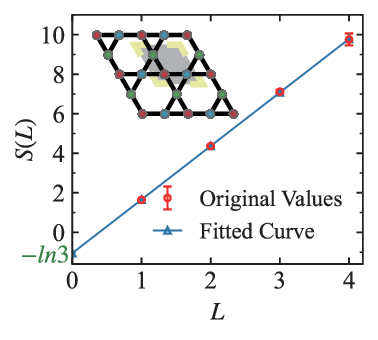}
    \put(-115, 85){(a)}
    \label{fig:entropy_z3}
    \end{minipage}
    }
    \subfigure{
    \begin{minipage}{0.47\linewidth}
    \centering
    \includegraphics[width = \linewidth]{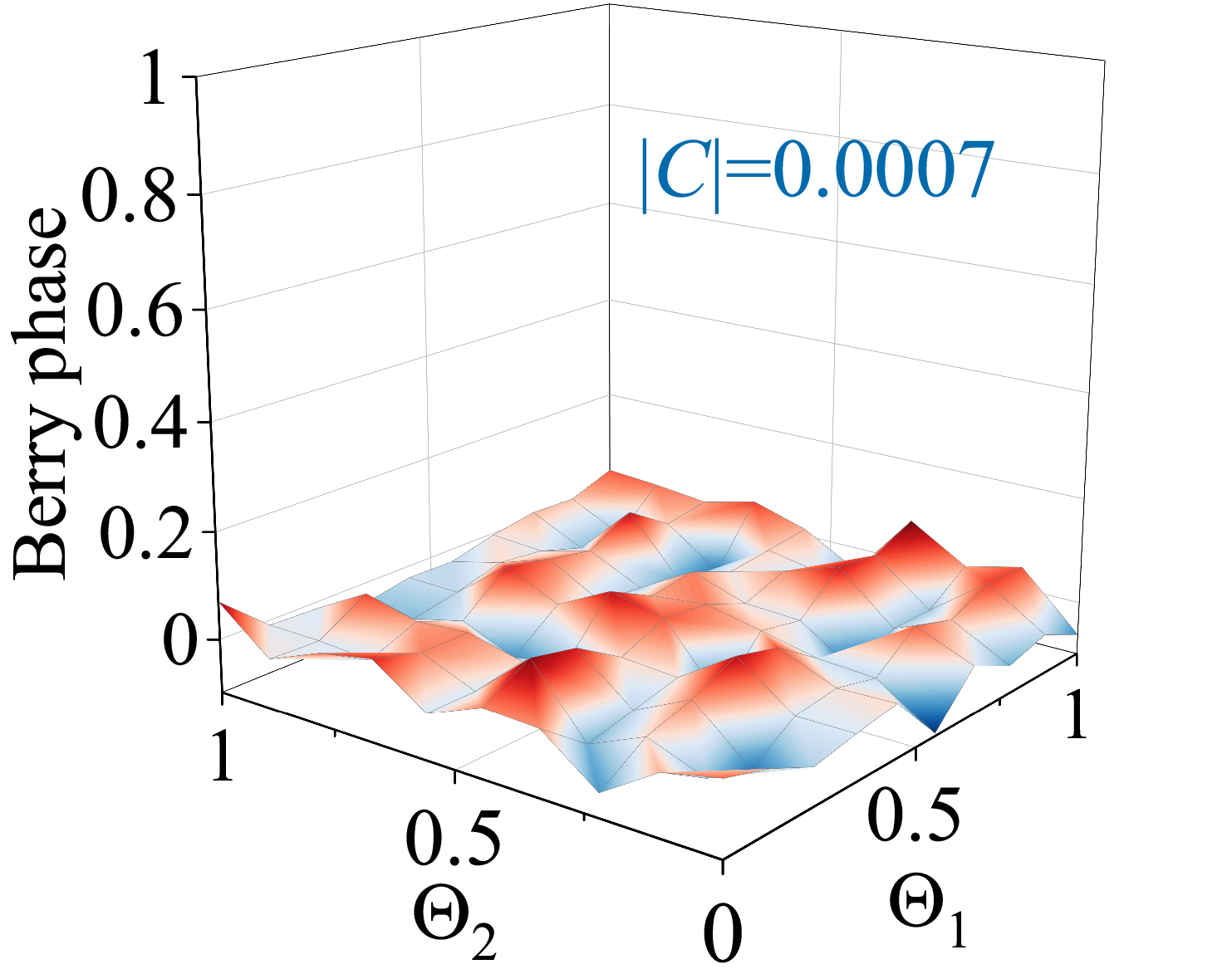}
    \put(-118,82){(b)}
    \label{fig:berry_phase_z3}
    \end{minipage}
    }
    \subfigure{
    \begin{minipage}{0.47\linewidth}
    \centering
    \includegraphics[width = \linewidth]{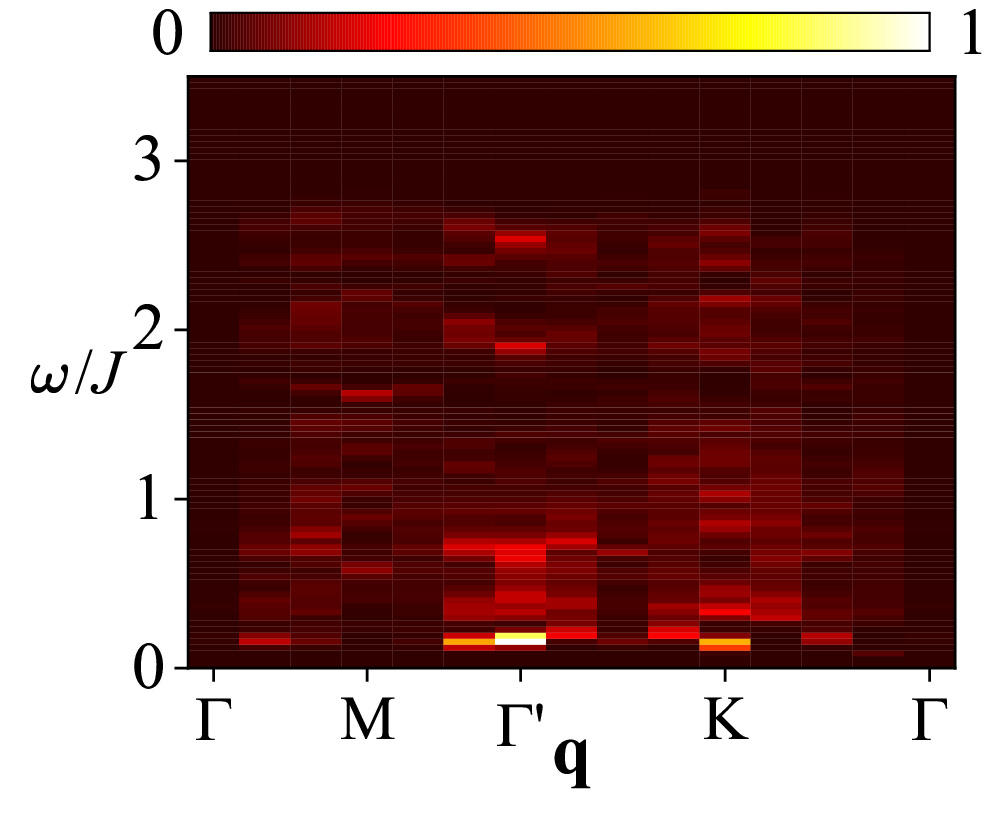}
    \put(-118,83){(c)}
    \label{fig:z3_sqw}
    \end{minipage}
    }
    \subfigure{
    \begin{minipage}{0.45\linewidth}
    \centering
    \includegraphics[width = \linewidth]{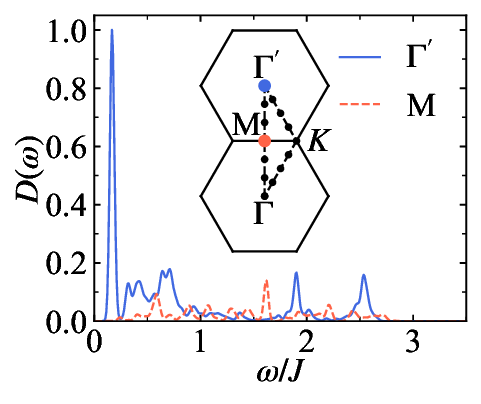}
    \put(-115, 80){(d)}
    \label{fig:z3_sw_g2_m}
    \end{minipage}
    }

    \caption{(a) Entanglement entropy of the 1/9 magnetization plateau phase as a function of subsystem size.  We choose the shape of the subsystem as a diamond, as indicated by the shaded area in the inset. The horizontal axis means that the area of the subsystem is $L^2$ times of primitive cell. The best fit to $S(L) = \alpha L - \gamma$ gives  $\gamma \sim 1.05$. (b) Distribution of Berry phase in the $\Theta_{1}$-$\Theta_{2}$ space, which is discretized into a grid of 100 plaquettes.  The Berry phase over each small plaquette is approximately proportional  to the Berry curvature. (c) Spin excitation spectrum for a lattice size of $6\times6\times3$. (d) Energy distribution curves of the spectra at $\Gamma^{\prime}$ and M. The insert shows the momentum path $\Gamma$-M-$\Gamma^\prime$-K-$\Gamma$ used in (c).}
    \label{fig:entropy_and_berry_phase_sqw}
\end{figure}

The ground-state energy curve depicted  in Fig.~\ref{fig:en_m_chi_vs_bz}(a) clearly shows two
phase transitions: one from DSL to $\mathbb{Z}_3$ QSL at $B_z/J \approx 0.35$, the other from $\mathbb{Z}_3$ QSL to a $\sqrt{3}\times\sqrt{3}$ VBS~\cite{supple} at $B_z/J \approx 0.63$. In Fig.~\ref{fig:en_m_chi_vs_bz}(b), the magnetization ratio $M / M_s$ of the field-induced $\mathbb{Z}_3$ QSL for $0.35 \alt B_z/J \alt 0.63$  is observed to stabilize at 1/9. This forms a robust magnetization plateau, which aligns with the experimental findings in compounds YCu$_3$(OH)$_{6+x}$Br$_{3-x}$ and YCu$_3$(OD)$_{6+x}$Br$_{3-x}$~\cite{arxiv.2310.07989,arxiv.2310.10069,arxiv.2311.11619}. From the perspective of spinons, the first five of the nine available spinon bands [see Fig.~\ref{fig:ek_z3}] are occupied by spin-up spinons, while spin-down spinons occupy only the first four, resulting in a magnetization ratio of $M / M_s = 2(N_{\uparrow} - N_{\downarrow}) / N = 1/9$. We also notice that the $\sqrt{3}\times\sqrt{3}$ VBS for $B_z/J > 0.63$ exhibits a 1/3 magnetization plateau, consistent with the experimental observations~\cite{arxiv.2310.07989,arxiv.2310.10069}. Furthermore, we calculate the chiral order parameter defined as $|\chi|=|\sum_{i \in \bigtriangleup/\bigtriangledown}\boldsymbol{S}_{i1}\cdot(\boldsymbol{S}_{i2}\times\boldsymbol{S}_{i3})|/N_{\bigtriangleup/\bigtriangledown}$~\cite{Wen-gsd-csl-1989, PhysRevB.92.125122}, where the indices 1, 2, and 3 correspond to vertexes (arranged clockwise) in an elementary triangle $i$, and $N_{\bigtriangleup/\bigtriangledown}$ represents the total number of triangles. As depicted in Fig.~\ref{fig:en_m_chi_vs_bz}(c), $|\chi|$ is found to be non-zero and constant throughout the 1/9 plateau phase, while it is zero in the DSL and $\sqrt{3}\times\sqrt{3}$ VBS phases. In the magnetization process, the only nonzero chirality of this $\mathbb{Z}_3$ QSL can be detected experimentally using polarized neutron scattering~\cite{PhysRevX.12.021029}.

We then examine the distribution of magnetization across the lattice for the 1/9 magnetization plateau phase. As shown in Fig.~\ref{fig:m_persite}, the magnetization $M / M_s$ at each site is very close to 1/9. Such uniformity is in stark contrast to the behavior seen in the magnetization plateau phases of other frustrated antiferromagnets~\cite{PhysRevB.67.104431,PhysRevB.76.060406,PhysRevLett.108.057205,PhysRevLett.109.267206,PhysRevLett.109.257205,PhysRevLett.110.267201,
nat.com.9.2666,JPSJ.65.2374,nat.phys.19.1883}, which typically exhibit a non-uniform magnetization within their expanded unit cells. On the other hand, though the magnetic moment distribution superficially resembles that of conventional ferromagnetic states, the underlying spin-spin correlations are profoundly different. As depicted in Fig.~\ref{fig:spin_cor}, the equal-time spatial spin-spin correlation function $C(r)$ in this phase exhibits a rapid decay to zero with the distance between pairs of sites. This behavior differs essentially from the long-range correlations of ferromagnetic order, where spins at infinitely separated distances remain perfectly correlated. The presence of such short-range spin-spin correlations is a distinguishing characteristic of a QSL.

\begin{table}
    \centering
    \caption{Eigenvalues of the $9\times9$ overlap matrix with the elements $\mathcal{O}_{\alpha\beta;\alpha'\beta'} = \langle \Psi_{\alpha, \beta} | \Psi_{\alpha',\beta'} \rangle$ for the $\mathbb{Z}_3$ QSL, calculated with lattice size $N = 12 \times 12 \times 3 = 432$.}
    \begin{tabular}{c|c|c|c|c|c|c|c|c}
        \hline \hline
        $\varepsilon_1$ & $\varepsilon_2$ & $\varepsilon_3$ & $\varepsilon_4$ & $\varepsilon_5$ & $\varepsilon_6$ & $\varepsilon_7$ & $\varepsilon_8$ & $\varepsilon_9$ \\
        \hline
        3.547 & \ 0.916  & \ 0.901 & \ 0.875 & \ 0.837 & \ 0.825 & \ 0.809 & 0.146 & 0.144 \\
        \hline \hline
    \end{tabular}
    \label{tab:gsd}
\end{table}

Our subsequent analysis focuses on the topological properties of the 1/9 magnetization plateau phase. An important quantity for characterizing topological properties is the TEE~\cite{PhysRevLett.96.110405,PhysRevLett.96.110404,PhysRevLett.104.157201, PhysRevLett.107.067202,nat.phys.8.902,PhysRevB.84.075128}. To obtain the TEE, we partition the system into two subsystems A and B, and calculate the Renyi entropy $S_{n}=(1-n)^{-1}\log[\mathrm{Tr}(\rho_{A}^{n})]$, where $\rho_{A}=\mathrm{Tr}_{B}|\Psi\rangle\langle\Psi|$ and $|\Psi\rangle$ is the ground-state wave function~\cite{supple,PhysRevB.84.075128}. For a short-ranged Hamiltonian, the entanglement entropy is predicted to follow $S(L) = \alpha L - \gamma$, with $L$ representing the boundary length of a contractible patch with codimension-1 boundary. The coefficient $\alpha$ is $n$-dependent, while $\gamma$, the TEE, is independent of $n$. We focus on the Renyi entropy with index $n = 2$, which is more feasibly to compute with our VMC method~\cite{supple,PhysRevB.84.075128}. Moreover, this TEE can reflect the total quantum dimension $D$ of the topological order, i.e., $\gamma = \ln D$. To extract the TEE, we calculate the entropy $S_{2}$ for varying sizes of the shaded region and apply a linear extrapolation to $L\rightarrow0$ in order to eliminate the area-law-associated $\alpha L$ term. As shown in Fig.~\ref{fig:entropy_z3}, the TEE of the 1/9 magnetization plateau phase is $\gamma \approx 1.05 \approx \ln 3$, suggesting that the total quantum dimension $D$ should be 3. Therefore, we can infer that this disordered phase with 1/9 magnetization is a $\mathbb{Z}_3$ QSL.

The nontrivial topological nature of this chiral $\mathbb{Z}_3$ QSL can be further characterized by its GSD~\cite{PhysRevB.40.7387,PhysRevB.41.9377, toric_kitaev}. We have constructed 9 projected ground states by applying different boundary conditions to the mean-field Hamiltonian $H_{\mathrm{mf}}$, each corresponding to varying magnetic fluxes threading the two hole of the torus lattice~\cite{supple}. These states are denoted as $|\Psi_{\alpha,\beta} \rangle = P_G |GS_{\alpha,\beta} \rangle_{\mathrm{mf}}$, where the fluxes $\alpha$ and $\beta$ take on the values of $2n\pi/3$ with $n \in \{0,1,2\}$. The ground-state degeneracy aligns with the linear independence of these 9 variational wave functions. To elucidate this degeneracy, we computed the overlaps between each pair of the nine states, assembling an overlap matrix~\cite{supple,PhysRevB.91.041124,PhysRevB.94.075131}. Analysis of this matrix revealed that all nine of its eigenvalues are nonzero, confirming that the GSD is $n_g = 9$, as summarized in Table~\ref{tab:gsd}. Given that the total quantum dimension is $D^2=9$, it shows that this chiral $\mathbb{Z}_3$ QSL manifests an Abelian topological order, satisfying the relation $D^2 = n_g$.

Moreover, unlike other chiral spin liquids~\cite{PhysRevB.91.041124,PhysRevB.94.075131,PhysRevB.52.4223}, the chiral $\mathbb{Z}_3$ QSL discussed here has a topological Chern number of zero. As shown in Fig.~\ref{fig:ek_z3}, the mean-field spinon dispersion exhibits gaps between any two bands, thereby the Chern number of each band is well-defined (see Table~\ref{tab:chern_num_z3}). From the perspective of spinons, the Chern number arising from the spin-up spinons is $C_{\uparrow} = \sum_{i=1}^5 C_{i} = -1$ for the magnetization ratio $M/M_s = 1/9$, while the spin-down spinons yield a Chern number $C_{\downarrow} = \sum_{i=1}^4 C_{i} = 1$. Thus, the total Chern number is zero, but the spin Chern number $C_{s}=(C_{\uparrow} - C_{\downarrow})/2$ is nonzero, which is similar to that of quantum spin Hall states~\cite{PhysRevLett.97.036808}. To verify the zero Chern number beyond the mean-field level, we construct the projective many-body wave functions with twisted boundary condition: $c_{i+L_{j},\uparrow} = c_{i,\uparrow} e^{i\Theta_{j}}$ and $c_{i+L_{j},\downarrow} = c_{i,\downarrow} e^{-i\Theta_{j}}$, with $j=1,2$ indicating the two primitive lattice vector directions, $L_{j}$ the lattice size along the $j$ direction and $\Theta_{j} \in [0, 2\pi]$ the twisted boundary phase. The Chern number is calculated by integrating the Berry curvature $\mathcal{F}(\Theta_{1},\Theta_{2})$~\cite{supple,PhysRevB.91.041124,PhysRevB.31.3372}:
$C=\frac{1}{2\pi}\int_{0}^{2\pi} d\Theta_{1}\int_{0}^{2\pi}d\Theta_{2}\mathcal{F}(\Theta_{1},\Theta_{2})$.
As  depicted in Fig.~\ref{fig:berry_phase_z3}, the Berry curvature has both positive and negative values, resulting in a net Chern number of zero. This zero Chern number also implies a zero chiral central charge~\cite{10.1093/nsr/nwv077}. Considering its long-range entanglement and the GSD of $n_{g}=9$,  we can identify this $\mathbb{Z}_3$ QSL as a topologically ordered phase with a rank 9 topological order, denoted as $9_0^B$ in Ref.~\onlinecite{10.1093/nsr/nwv077}. Moreover, the non-zero chirality of this $\mathbb{Z}_3$ QSL is also consistent with the existence of two $9_0^B$ states that break time-reversal symmetry and are mutual time-reversed states.

\begin{table}
    \centering
    \caption{Chern numbers $C$ of the mean-field spinon bands for the $\mathbb{Z}_3$ QSL.}
    \begin{tabular}{c|c|c|c|c|c|c|c|c|c}
        \hline \hline
        Index & \ \ \ 1\ \ \ & \ \ \ 2\ \ \ & \ \ \ 3\ \ \ & \ \ \ 4\ \ \ & \ \ \ 5\ \ \ & \ \ \ 6\ \ \ & \ \ \ 7\ \ \ & \ \ \ 8\ \ \ & \ \ \ 9\ \ \ \\
        \hline
        $C$ & 1 & -2 & 1 & 1 & -2 & 4 & -2 & -2 & 1 \\
        \hline \hline
    \end{tabular}
    \label{tab:chern_num_z3}
\end{table}

Finally, we discuss how to experimentally identify the $\mathbb{Z}_3$ QSL by measuring the spin excitation spectrum. Figure~\ref{fig:z3_sqw} shows the longitudinal dynamic structure factor $D(\boldsymbol{q},\omega)$~\cite{supple} calculated using the VMC method~\cite{PhysRevB.97.235103, PhysRevB.102.195106, 10.21468/SciPostPhys.14.6.139}. The excitation spectrum is gapped and manifests as a broad continuum, originating from the fractionalization of the $S=1$ spin excitations. A notable feature is the enhanced periodicity of its lower edge, as evidenced by the presence of multiple minima with the same energy in the first BZ. This is related to the translation symmetry fractionalization~\cite{PhysRevB.87.104406,PhysRevB.97.115138}, and the unique fractionalization characteristics of the $\mathbb{Z}_3$ QSL can be used to distinguish it from other QSL states~\cite{PhysRevB.102.195106, 10.21468/SciPostPhys.14.6.139}. Additionally, the $\omega$-dependence of the spectra at specific momenta can also serve as a basis for experimental identification of the $\mathbb{Z}_3$ QSL. As shown in Fig.~\ref{fig:z3_sw_g2_m}, the spectra at $\Gamma^{\prime}$ and M exhibit several characteristic peaks, and they differ significantly from the spectra of other QSL states in the kagome system~\cite{PhysRevB.102.195106, 10.21468/SciPostPhys.14.6.139}.

In summary, motivated by experimental observations of the 1/9 magnetization plateaus in YCu$_3$(OH)$_{6+x}$Br$_{3-x}$ and YCu$_3$(OD)$_{6+x}$Br$_{3-x}$, we utilize the VMC method to investigate the magnetization of the antiferromagnetic Heisenberg model on the kagome lattice, with a particular emphasis on elucidating the nature of the 1/9 magnetization plateau. By increasing the magnetic field, we observe a field-induced magnetically disordered phase exhibiting a 1/9 magnetization plateau. Detailed investigations of the magnetization pattern, spin correlations, chiral order parameter, and topological entanglement entropy have led us to identify this 1/9 magnetization plateau phase as a chiral $\mathbb{Z}_3$ topological QSL. We also highlight key features in the spin excitation spectrum that can be used for the experimental identification of this $\mathbb{Z}_3$ QSL. It should be noted, however, that our model does not include disorder effects, which are unavoidable in real materials. The influence of disorder effects on the magnetization plateau phase is also an important issue that warrants further study.
\begin{acknowledgments}
We thank Q.-H. Wang, Z.-X. Liu and Z.-L. Gu for help discussions. This work was supported by National Key Projects for Research and Development of China (Grant No. 2021YFA1400400) and the National Natural Science Foundation of China (No. 92165205, No. 12074175 and No. 12374137).
\end{acknowledgments}

\bibliography{ref}

\end{document}


\title{Supplemental Material for ``Variational Monte Carlo Study of the 1/9 Magnetization Plateau in Kagome Antiferromagnets"}

\author{Li-Wei He}
\affiliation{%
 National Laboratory of Solid State Microstructures and School of Physics, Nanjing University, Nanjing 210093, China
}%

\author{Shun-Li Yu}%
\email{slyu@nju.edu.cn}
\affiliation{%
 National Laboratory of Solid State Microstructures and School of Physics, Nanjing University, Nanjing 210093, China
}%
\affiliation{Collaborative Innovation Center of Advanced Microstructures, Nanjing University, Nanjing 210093, China}

\author{Jian-Xin Li}%
\email{jxli@nju.edu.cn}
\affiliation{%
 National Laboratory of Solid State Microstructures and School of Physics, Nanjing University, Nanjing 210093, China
}%
\affiliation{Collaborative Innovation Center of Advanced Microstructures, Nanjing University, Nanjing 210093, China}

\maketitle

\section{\label{sec:mf_ansatz}MEAN-FIELD DECOUPLING}
The antiferromagnetic (AFM) Heisenberg spin model we consider in the main text can be rewritten by fermionic doublet representation as the following,
\begin{equation}
    \begin{aligned}
        &\boldsymbol{S}_i \cdot \boldsymbol{S}_j = -\frac{1}{4}(T_{ij}T_{ij}^\dagger + P_{ij}P_{ij}^\dagger) + \mathrm{const},
    \end{aligned}
\label{eq:fermionic_representation}
\end{equation}
where $T_{ij} = \psi^\dagger_i \psi_j$ ($P_{ij} = \psi^\dagger_i \bar{\psi}_j$) is the singlet hopping (pairing) term with $\psi = (c_{\uparrow}, c_{\downarrow})^T$, $\bar{\psi} = (c_{\downarrow}^\dagger, -c_{\uparrow}^\dagger)$. Because of the SU(2) gauge structure of this fermionic representation, it is necessary to implement Lagrangian multipliers $\boldsymbol{\lambda}$ to constraint condition, namely enforcing generators of SU(2) gauge group $\boldsymbol{\Lambda}_i = 0$ to return to the subspace of real physical state in the mean-field theory~\cite{Zhao-Liu-j4-prl-2021}. Their expression with fermionic doublet representation as following,
\begin{equation}
    \begin{aligned}
        &\Lambda_i^x = -\frac{1}{4}(\psi_i^\dagger \bar{\psi}_i + \bar{\psi}_i^\dagger \psi_i),\\
        &\Lambda_i^y = -\frac{i}{4}(\psi_i^\dagger \bar{\psi}_i - \bar{\psi}_i^\dagger \psi_i),\\
        &\Lambda_i^z = \frac{1}{2}(1 - \psi_i^\dagger \psi_i).
    \end{aligned}
\end{equation}
One can notice that the $\lambda_z$ term is just equivalent to the local particle number constraint while the $\lambda_x$ and $\lambda_y$ terms compose into the on-site pairing constraint that is the deduction of the former one~\cite{Wen-psg-2002}.

When the external magnetic field $B_z$ is turned on, the model still preserves the U(1) spin rotational symmetry around the spin-$z$ direction. Namely, the $S_z$ is always conserved, consequently the role of the field is to tune the chemical potential. Therefore, we can decouple spin interactions into noninteracting quadratic structure to obtain the following full mean filed Hamiltonian (we omit an irrelevant constant)
\begin{equation}
    \begin{aligned}
    H_{\mathrm{mf}} =& \sum_{i,j}( t_{ij} \psi_i^\dagger \psi_j +
    \Delta_{ij} \psi_i^\dagger \bar{\psi}_j + \mathrm{H.c.}) \\
    &+ \sum_{i} \boldsymbol{\lambda} \cdot \boldsymbol{\Lambda}_i
    - \mu \psi^\dagger_i  \sigma_z \psi_i,
    \end{aligned}
\label{eq:full-mf-hamiltonian}
\end{equation}
where $t_{ij}$ ($\Delta_{ij}$) is the spinon hopping (pairing) parameter, and $\mu$ is the chemical potential tuned by the field. Therefore, all the variational parameters are $p = (t_{ij}, \Delta_{ij}, \boldsymbol{\Delta}^t_{ij}, \boldsymbol{\lambda}, \mu)$. Obviously, there must be various gauge non-equivalent ans\"atzs in Eq. (\ref{eq:full-mf-hamiltonian}) with so plenty of variational parameters. We selectively consider some of them according to the projective symmetry group~\cite{Wen-psg-2002, lu-2011-psg, Bieri-psg-tri-2016} (PSG) as shown below.

\section{\label{ansatz} DETAILS OF VARIOUS ANS\"ATZES}

\textit{uRVB}.---The uniform RVB state is considered because it is easiest to construct, i.e., all the first nearest-neighbor (NN) hopping terms $t_{ij} = 1.0$ and the rest are vanishing.

\textit{DSL}.---Now let's consider one of the most competitive candidate ground states in the AFM $J_1$ Heisenberg model on kagome lattice when the field is absent, Dirac spin liquid (DSL). In this DSL, if we only consider the first nearest-neighbor (NN) hopping ones, there are zero ($\pi$) fluxes through triangles (hexagons), respectively. From PSG analysis~\cite{lu-2011-psg, Bieri-psg-tri-2016}, the second singlet hopping terms can also exist in principle while the third NN terms in this state are forbidden by PSG so that we rationally abandon them. However, we find the second hopping term almost disappears according to the VMC calculations.

\textit{Chiral QSL}.---The chiral spin liquid (CSL) is the topological order with the $\theta$ ($\pi - 2\theta$) fluxes through triangles (hexagons). Actually, the above DSL is a particular case of this CSL for $\theta = 0$. This state, supporting the semionic excitation, is the lattice version of the $\nu = 1/2$ Kalmeyar-Laughlin state and could be stabilized by the third NN AFM Heisenberg $J_3$ and three-spin chiral interaction~\cite{hu-j12d_ch-prb-2015-csl}.

\begin{figure}
	\centering
	\subfigure[]{
		\includegraphics[width = 0.4\linewidth]{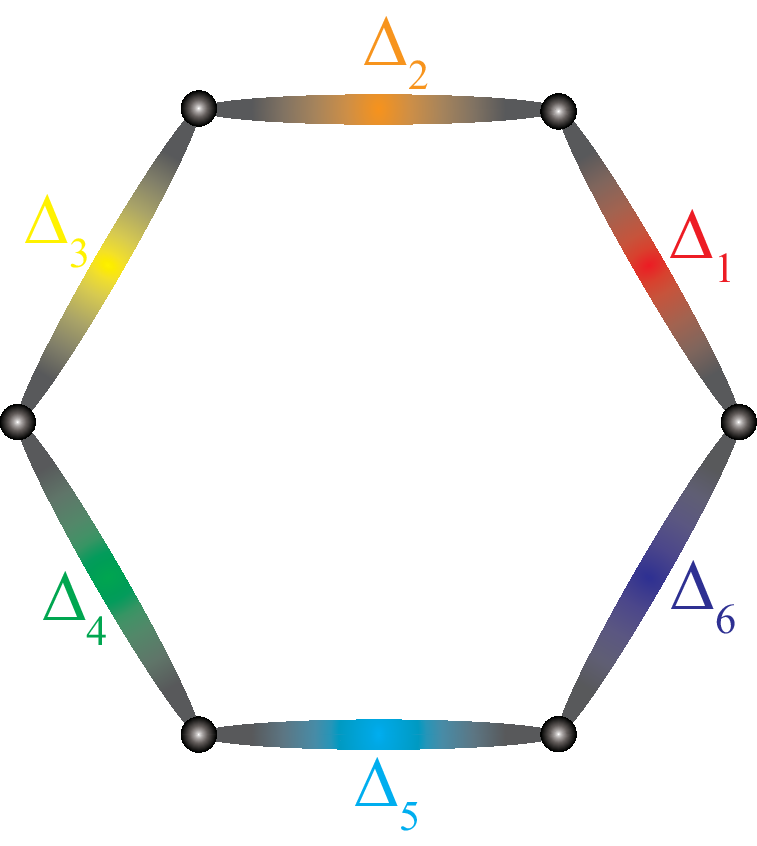}
		\label{fig:pairing}
	}
     \subfigure[]{
		\includegraphics[width = 0.4\linewidth]{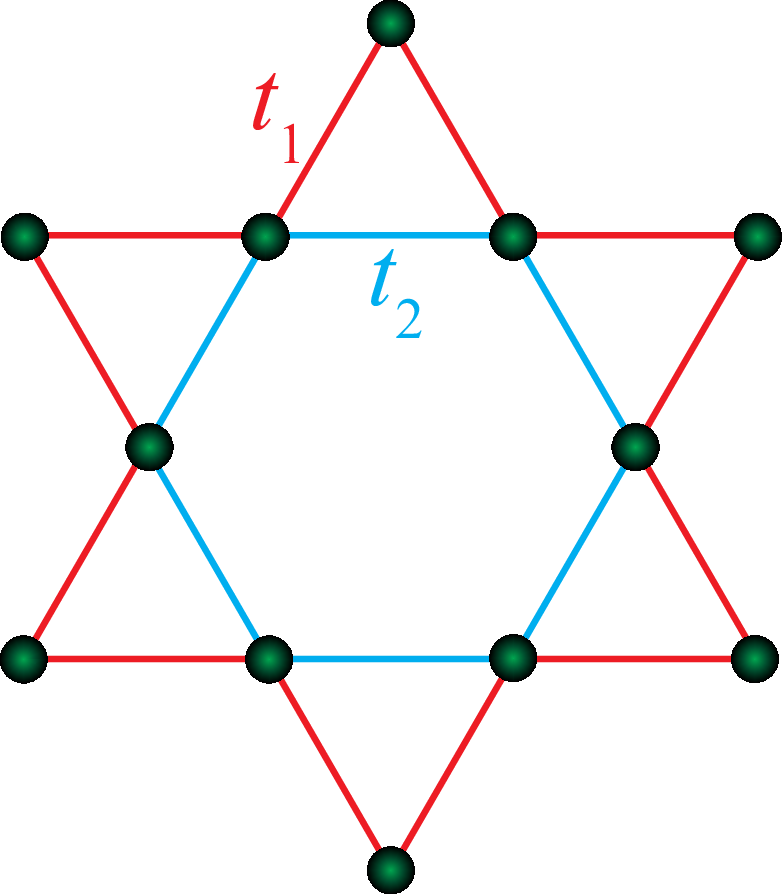}
		\label{fig:vbs}
	}%
	\caption{(a) is the illustration of the pairing pattern for each hexagon on kagome lattice. In detail, $\Delta_{n} = \Delta e^{i \mathcal{L} n\pi/3}$, where $\mathcal{L} = \{0,1,2,3\}$ are the angular momenta for the $s$-, $p+ip$-, $d+id$- and $f$-wave spinon pairings, respectively, and $\pi/3$ is the azimuth angle of the first NN bonds. (b) is the hopping pattern of the David-star-type VBS state, $t_1$ is the hopping term on the twelve red bonds on the boundary of David star while $t_2$ is the one on the six blue bonds on the hexagon.}
	\label{fig:vbs_pairing}
\end{figure}

\textit{Z$_2$ QSL with spinon pairing}.---For this kind of QSLs, their invariant gauge group (IGG) is Z$_2$. According to PSG~\cite{lu-2011-psg}, a so-called Z$_2[0,\pi]\beta$ state is found, where $[0, \pi]$ just means the flux pattern is the same as the DSL. In fact, this Z$_2$ QSL is an adjoining phase of the DSL with the second NN spinon-pairing instability because the PSG of the two states are the same~\cite{Bieri-psg-tri-2016}. However, this gapped state is not energetically favored for AFM Heisenberg model~\cite{Iqbal-prb-j1j2-dsl}. By our calculation, we also find the optimal pairing term is almost zero (as we know, the pairing parameters are never exactly equal to zero in the variational process), i.e., this state degenerates into the DSL, which is consistent with the result in the Ref.~\cite{Iqbal-prb-j1j2-dsl}. Even though it is claimed that the $d+id$-wave ($p+ip$-wave) spinon pairing is gauge equivalent to the $s$-wave ($f$-wave) one~\cite{Bieri-psg-tri-2016}, we nonetheless consider all of them without loss of generality. Their pairing pattern is shown in Fig.~\ref{fig:vbs_pairing}(a). The hopping terms for those of DSL and $\mathbb{Z}_3$ QSL are adopted. However, all the paired states are not energetically favored by our calculation, i.e., the optimal amplitude of spinon pairing is always close to zero. Therefore, in the main text, the mean-filed Hamiltonian includes no pairing term $\Delta_{ij}$ and $\boldsymbol{\lambda}$.

\begin{figure}
    \centering
    \subfigure[]{
    \includegraphics[width = \linewidth]{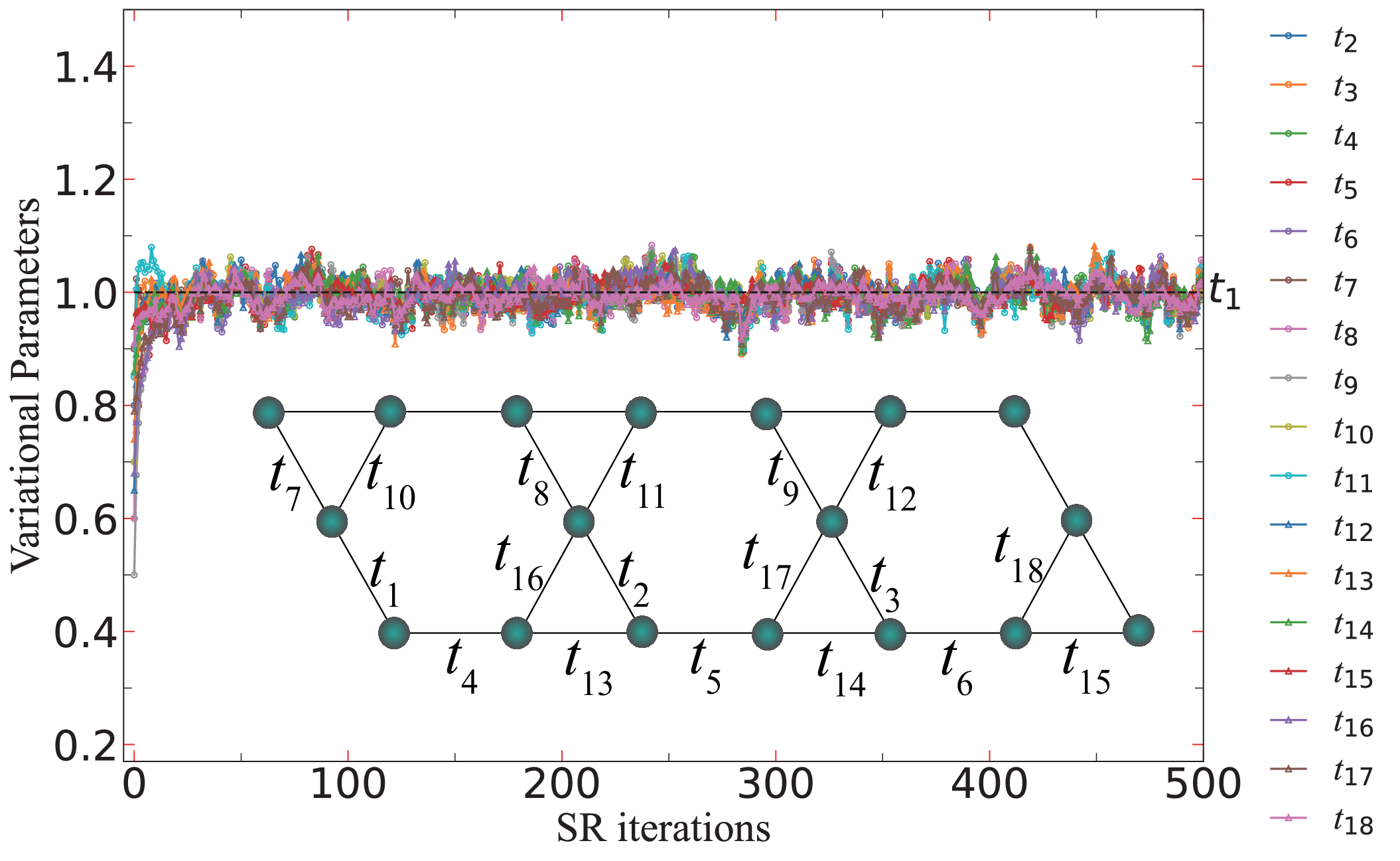}
    \label{fig:var_mod_z3_12_9}
    }
    \subfigure[]{
        \includegraphics[width = \linewidth]{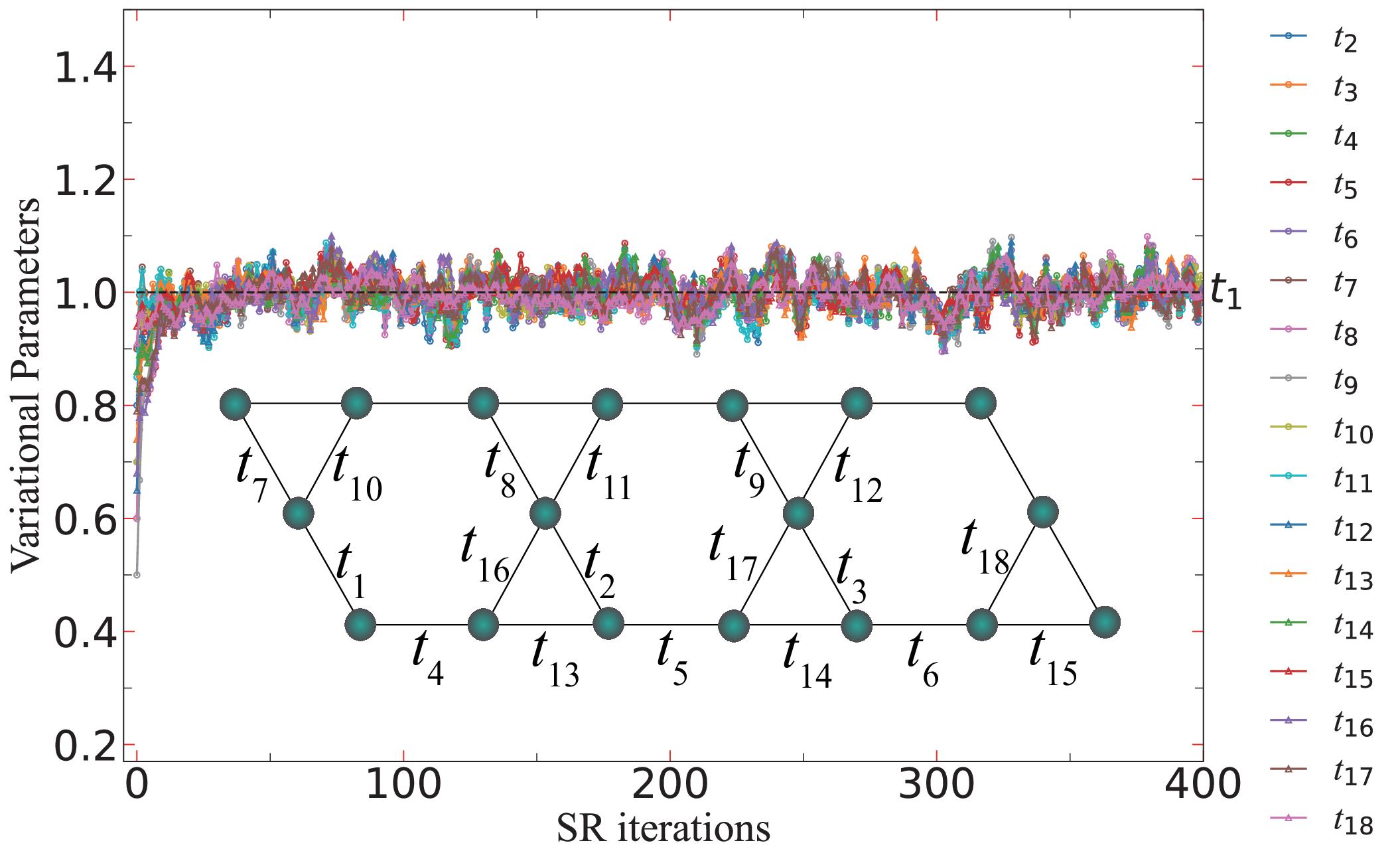}
        \label{fig:var_mod_z3_12_12}
    }
    \caption{(a) and (b) are the amplitudes of hopping terms as a function of SR iterations in VMC for the lattice sizes $12 \times 9 \times 3$ and $12 \times 12 \times 3$ at 1/9 magnetization plateau, respectively. The black dashed line means the $t_1 = 1.0$ as the reference and the remaining initial amplitudes are far from uniform. And the insert illustrates the distribution of 18 hopping terms.}
    \label{fig:var_mod_z3}
\end{figure}

\textit{David-star-type VBS state}.---As shown in Fig.~\ref{fig:vbs_pairing}(b), we construct the David-star-type VBS state~\cite{hasting_vbs_kagome}, whose unit-cell size is $\sqrt{3} \times \sqrt{3}$, as a possible instability of DSL. When the $|t_1| \neq |t_2|$, this situation gives the achiral masses. However, this situation increases the variational energy~\cite{Ran-dsl-2007}. Moreover, in our numerical analysis, we include the ratio $\delta = |t_1/t_2|$ in the variational parameter. We find that the optimal $\delta$ is almost 1, i.e., this VBS state is not energetically favored.

\textit{$\mathbb{Z}_3$ QSL}.---Its ans\"atz is elaborated in the main text. This exotic state triples the primitive cell with 18 different hopping terms $t_{ij} = e^{i\theta_{ij}}$ (we fix the amplitude of $t_{ij}$ to 1). The number of independent $\theta_{ij}$s is 15 because of a $2\pi/3$ flux through each primitive cell in the unit cell. To capture possible VBS state~\cite{kagome_plateau_prb_vbs} as an instability of this $\mathbb{Z}_3$ QSL at 1/9 magnetization plateau, we further include the 17 independent amplitudes (we can fix a amplitude to 1 as the reference in variational process) of hopping term in variational parameters and adopt the previously optimized phases $\theta_{ij}$s in the case of uniform amplitudes. However, by our calculation, we do not observe the possible VBS state, i.e., all the amplitudes are almost equal to 1 at this 1/9 plateau, as shown in Fig.~\ref{fig:var_mod_z3}.

\begin{figure}
    \centering
    \includegraphics[width=\linewidth]{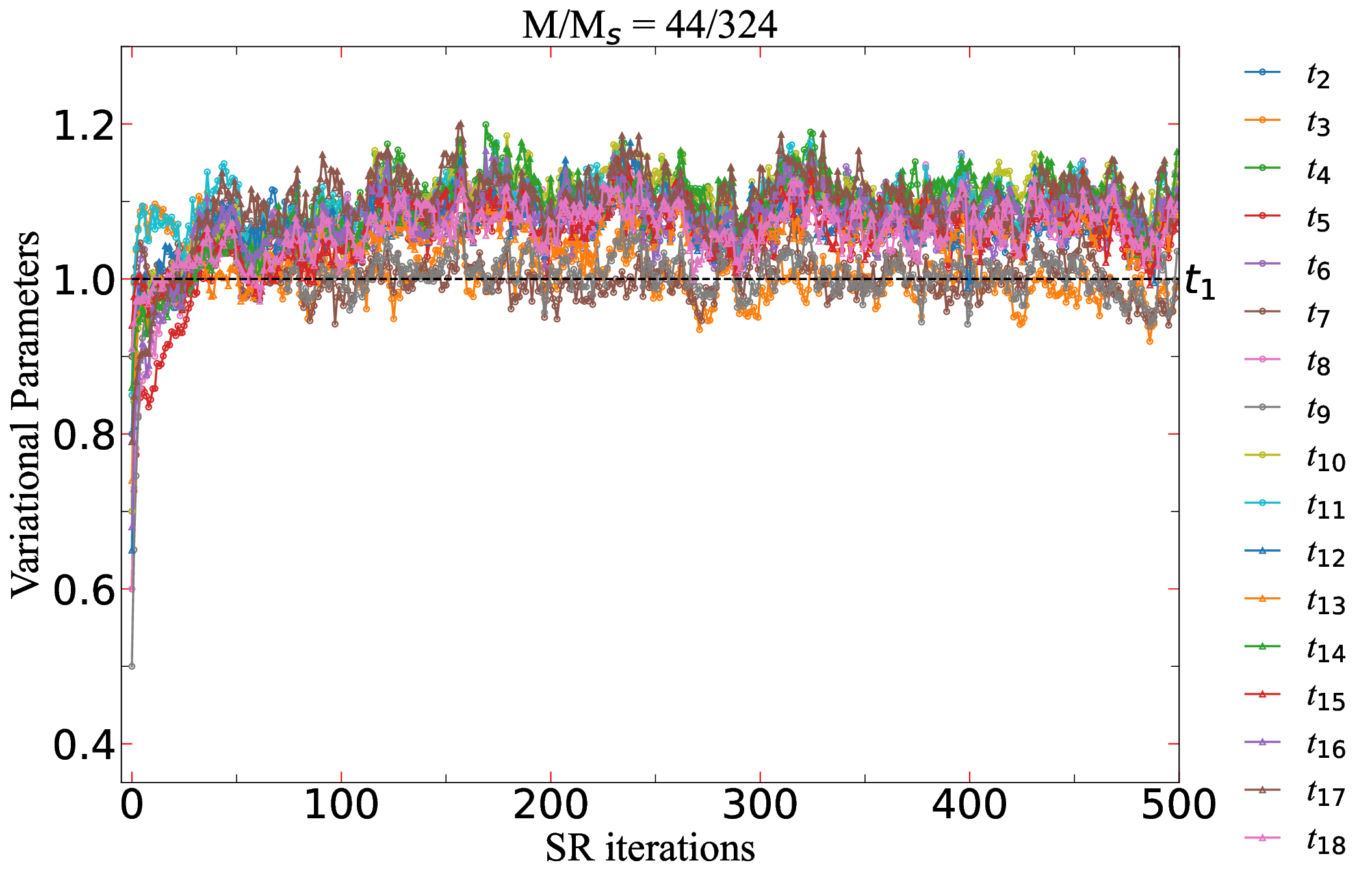}
    \caption{Non-uniform amplitudes of hopping terms as a function of SR iterations in VMC for the lattice size $12 \times 9 \times 3$ when the $M/M_s = 44 / 324 > 1/9$. And the distribution of hopping terms is the same as the insert of Fig.~\ref{fig:var_mod_z3}.}
    \label{fig:var_mod_vbs}
\end{figure}

\textit{VBS state based on the $\mathbb{Z}_3$ QSL}.---With the increase of the magnetization, we repeat the above variational calculation to search the possible VBS states with 9 kinds of sublattices as an instability of the $\mathbb{Z}_3$ QSL. To capture as much accurate information of this VBS state as possible, we iterate the above variational calculations. In detail, we apply the optimal amplitudes obtained in the previous iteration to optimize the phases, then, apply this optimal phases to optimize the amplitudes in the next iteration. The reason why we do this is because the number of variational parameters is so large. Besides, to further stabilize this numerical calculation, we reweight the last two optimal amplitudes (phases) to obtain a new one as the inputs in the current iteration to optimize phases (amplitudes). Specifically, in the $n$th iteration to optimize amplitudes, the new phases are $\theta_{ij, new} = \rho \theta_{ij, n - 2} + (1 - \rho) \theta_{ij, n - 1}$, where $\rho = 0.3$ we adopt is weight factor. After several iterations for different lattice sizes, we find the optimal phases almost retain unchanged, so we can regard the amplitudes and phases as independent of each other. Interestingly, we do find a possible VBS instability when $M/M_s > 1/9$, as shown in Fig.~\ref{fig:var_mod_vbs} and listed in Table.~\ref{tab:en_z3_vbs_vs_m} and \ref{tab:en_vbs_sqrt3_vs_m}. However, after recovering the contribution of magnetic field, this kind of VBS is not energetically favored when $B_z / J \agt 0.63$ but the following VBS state is dominant.

\begin{figure}[h]
    \centering
    \includegraphics[width=0.9\linewidth]{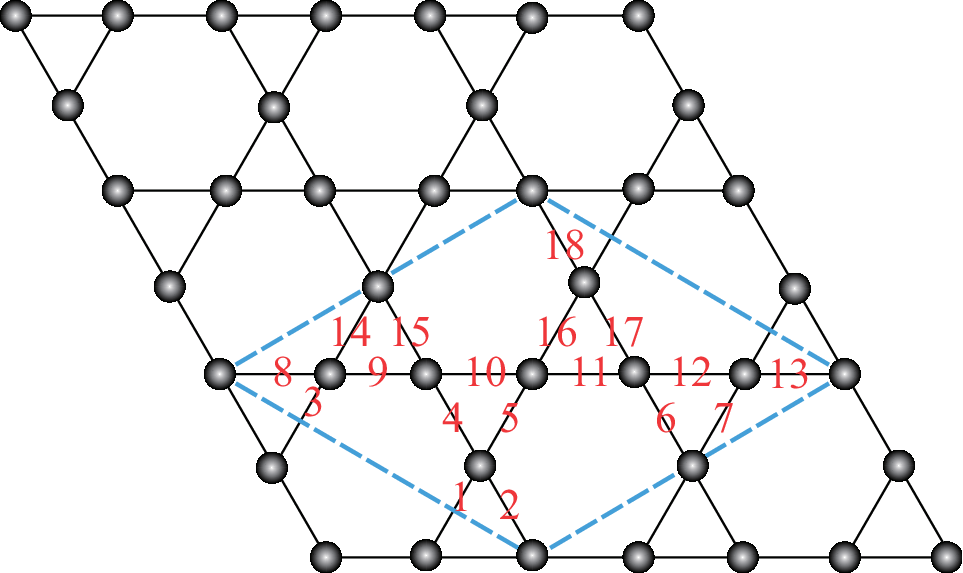}
    \caption{The blue dashed line means the unit cell of the $\sqrt{3} \times \sqrt{3}$ VBS state. The flux pattern is the same as $\mathbb{Z}_3$ QSL, namely, the flux through each primitive cell is $2\pi/3$ and parameterized to each NN bond carefully. The amplitudes of the 18 variational hopping parameters in the VMC are labeled by red numbers. The optimal variational amplitudes obtained by the VMC method in a $12 \times 12 \times 3$ system are: $t_1 = 1.073$, $t_2 = 1.059$, $t_3 = 1.027$, $t_4 = 1.075$, $t_5 = 1.070$, $t_6 = 1.016$, $t_7 = 0.909$, $t_8 = 1.026$, $t_9 = 1.014$, $t_{10} = 0.939$, $t_{11} = 1.022$, $t_{12} = 1.027$, $t_{13} = -0.011$, $t_{14} = 1.012$, $t_{15} = -0.005$, $t_{16} = -0.006$, $t_{17} = 1.000$, $t_{18} = 0.929$. Here, we fix $t_{17} = 1.000$ as the reference for all the other amplitudes. Besides, the reason of the minus sign of $t_{13, 15, 16} \sim 0$ is because we allow the amplitudes are negative numbers in our variational calculations.}
    \label{fig:vbs_sqrt3_ansatz}
\end{figure}

\textit{$\sqrt{3} \times \sqrt{3}$ VBS state stabilized in the $1/3$ magnetization plateau phase}.---This VBS state is the one in the $1/3$ magnetization plateau mentioned in the main text when $B_z / J \agt 0.63$. Its optimal variational parameters $t_{ij}$ are shown in Fig.~\ref{fig:vbs_sqrt3_ansatz}.

\begin{table}[ht]
    \centering
    \caption{Variational energy per site omitted the contribution from magnetic field as a function of magnetization for the lattice size $12 \times 12 \times 3$. The second (third) row labeled by $E_{1(2)}$ is the energy of the $\mathbb{Z}_3$ QSL (the VBS state based on the $\mathbb{Z}_3$ QSL) and the fourth one is that of $\sqrt{3} \times \sqrt{3}$ VBS state stabilized in the $1/3$ magnetization plateau phase. The error bars for all the energies are almost $ \sim 10^{-5}$, and the symbol ``$\backslash$'' means we do not calculate the corresponding energy since the phase transition has happened.}
    \begin{tabular}{c|c|c|c|c|c|c}
        \hline \hline
        $M/M_s$ & 1/9 & 50/432 & 56/432 & 60/432 & 64/432 & 1/3\\
        \hline
        $E_1$ & -0.41178 & -0.41003 & -0.40459 & -0.40077 & -0.39680 & $\backslash$ \\
        \hline
        $E_2$ & -0.41178 & $\backslash$ & -0.40459 & -0.40086 & -0.39687 & -0.33784\\
        \hline
        $E_3$ & $\backslash$ & $\backslash$ & -0.40455 & -0.40058 & -0.39658 & -0.34106\\
        \hline \hline
    \end{tabular}
    \label{tab:en_z3_vbs_vs_m}
\end{table}

\begin{table}[h]
    \centering
    \caption{Variational energy per site omitted the contribution from magnetic field as a function of magnetization for the lattice size $12 \times 9 \times 3$. The second (third) row labeled by $E_{1(2)}$ the energy of the VBS state based on the $Z_3$ QSL ($\sqrt{3} \times \sqrt{3}$ VBS state). The error bars for all the energies are almost $ \sim 10^{-5}$.}
    \begin{tabular}{c|c|c|c|c|c}
        \hline \hline
        $M/M_s$ & 48/324 & 60/324 & 84/324 & 96/324 & 1/3\\
        \hline
        $E_1$ & -0.39698 & -0.38213 & -0.35837 & -0.34813 & -0.33787\\
        \hline
        $E_2$ & -0.39688 & -0.38200 & -0.35791 & -0.34757 & -0.34103\\
        \hline \hline
    \end{tabular}
    \label{tab:en_vbs_sqrt3_vs_m}
\end{table}

Finally, based on the above ans\"atzes, we can construct the trial wave function by applying the Gutzwiller projector to the ground state of the various different mean-field Hamiltonian (\ref{eq:full-mf-hamiltonian}), $|\Psi \rangle = P_G |GS_{\mathrm{mf}}\rangle$.

\section{\label{sec:magnetization} MAGNETIZATION IN VMC}

\subsection{\label{vmc_u1} States without spinon pairing}

For these states without spinon pairing, the parameters $\lambda_{x,y}$ are not considered any more.  For a system with $2N$ sites, the form of projective state can be expanded by the real-space configuration $|x\rangle$ as follows,
\begin{equation}
    |\Psi \rangle =  \sum_x \langle x | \Psi \rangle |x\rangle, \langle x | \Psi \rangle = \det(A),
    \label{eq:inner_prod_unpair}
\end{equation}
$A$ is a $2N \times 2N$ matrix with the elements $A_{ij} = \langle 0 |c_{i,\sigma_i} \phi_j^\dagger|0\rangle$, where $\phi_j$ is the $j$th eigenvector of the mean-field Hamiltonian. When $B_z = 0$, these states with only singlet hopping terms (up spinon and down spinon are not coupled to each other) should preserve $S_z = 0$, i.e., $N_\uparrow = N_\downarrow = N$. Thus, $\det(A) = \det(A_\uparrow) \det(A_\downarrow)$, where $A_{\uparrow(\downarrow)}$ is an $N_{\uparrow(\downarrow)} \times N_{\uparrow(\downarrow)}$ matrix related to the spin-up (spin-down) spinon mean-field Hamiltonian. When a finite field $B_z$ is turned on, the filling of spinons with opposite spins is tuned through the parameter $\mu$ at the mean-field level. In practical calculation for a finite system, we sweep all the sectors of total $S_z = (N_\uparrow - N_\downarrow)$ one by one and compare their variational energies to obtain the optimal $S_z$, corresponding to the magnetization for a finite field.

\subsection{\label{vmc_z2} States with spinon-singlet pairing}

For the states with spinon-singlet pairing as described in Sec.~\ref{sec:mf_ansatz}, the general ground state is the BCS-type wave function, $|GS_{\mathrm{mf}}\rangle = \prod_{ij,\sigma_i\sigma_j} (1 + a_{i\sigma_i,j\sigma_j}c_{i\sigma_i}^\dagger c_{j\sigma_j}^\dagger)|0\rangle$. For a system with $2N$ sites (to which the subsequent discussion is primarily dedicated, unless specified otherwise), the projective state is rewritten as follows,
\begin{equation}
    |\Psi \rangle = \sum_x \langle x | \Psi \rangle |x\rangle, \langle x | \Psi \rangle = \mathrm{Pf}(R),
    \label{eq:inner_prod_pair}
\end{equation}
where $R$ is a $2N \times 2N$ skew-symmetry matrix with elements $R_{ij} = a_{i\sigma_i,j\sigma_j}$. In our work, the spinon pairing is only singlet, so the form of $|GS_{\mathrm{mf}}\rangle$ can be reduced by $\sigma_i = \bar{\sigma}_j$. Meanwhile, the $R$ matrix is also reduced as follows,
\begin{equation}
    R =
    \begin{pmatrix}
    0 & B\\
    -B^T & 0\\
    \end{pmatrix},
\end{equation}
and the element of $N \times N$ matrix $B$ is $B_{ij} = a_{i\uparrow,j\downarrow}$. Then, one can notice that $\mathrm{Pf}(R) = \det(B)$. Finally, this kind of projective ground state in our work is composed of the superposition of various spinon-singlet configurations, i.e., RVB state with total $S_z = 0$.

The finite magnetization must happen when the field is large enough. It results in total $S_z = (N_\uparrow - N_\downarrow)/2 > 0$. In this case, for arbitrary configuration, it is impossible to construct $N$ pairs of spinon singlets. This means that the original $\det(B)$ is no longer valid. Now, for a certain total $S_z > 0$, $N_\uparrow - N_\downarrow = N_\delta > 0$, $N_\delta$ is the number of the unpaired spin-up spinons, while $N_\downarrow$ that of the singlets. In addition, there is no freedom to select which $N_\delta$ spin-up spinons remain unpaired, so all possible scenarios should be considered. Thus, the new inner production $\langle x | \Psi \rangle$ should be expressed as follows,
\begin{equation}
    \langle x | \Psi \rangle = \sum_{i} \det(A_i)\det(B_i),
\end{equation}
where the matrices $A_i$ and $B_i$, representing the inner products of the unpaired and paired spinon sectors, are of dimensions $N_\delta \times N_\delta$ and $N_\downarrow \times N_\downarrow$, respectively. The index $i$ denotes the various possible configurations specifying which $N_\delta$ spin-up spinons are unpaired, with the remainder forming pairs. The elements of $A_i$ ($B_i$) are similar to those of Eq.~\ref{eq:inner_prod_unpair} (\ref{eq:inner_prod_pair}). In fact, according to the property of the determinant in linear algebra, the superposition of all possible $\det(A_i)\det(B_i)$ is just a determinant of an expanded matrix, $\det(\widetilde{B}) = \langle x | \Psi \rangle$, which is composed of two matrices as follows,
\begin{equation}
    \widetilde{B} = (A', B'),
\end{equation}
where $A'$ ($B'$) is an $N_\uparrow \times N_\delta$ ($N_\uparrow \times N_\downarrow$) matrix. Now let's elaborate these two matrices. For a mean-field Hamiltonian $H_t$ without sipnon pairing, there must be $2N$ eigenvalues $E_n$ and the corresponding eigenvectors $\phi_n$, and we put them in order, $E_1 < E_2 < \ldots < E_n$. The ground state should be firstly filled by the first $N_\delta$ quasi-particles with the lowest energy. We can pack the corresponding $N_\delta$ eigenvectors into a matrix, $U = (\phi_1, \ldots, \phi_{N_\delta})$. Assuming a real-space configuration $|x\rangle = c_{r_1 \uparrow}^\dagger \ldots c_{r_{N_\uparrow} \uparrow}^\dagger c_{r_{N_\uparrow + 1} \downarrow}^\dagger \ldots c_{r_{N} \downarrow}^\dagger |0\rangle$, the concrete forms of $A'$ are as follows,
\begin{equation}
    A' =
    \begin{pmatrix}
    U_{r_1, 1} & \ldots & U_{r_1, N_\delta}\\
    \vdots & \ddots & \vdots \\
    U_{r_{N_\uparrow}, 1} & \ldots & U_{r_{N_\uparrow}, N_\delta}\\
    \end{pmatrix}.
\end{equation}
When the spinon pairing term $H_\Delta$ is included, the elements of matrix $B'$ should be as follws,
\begin{equation}
    B' =
    \begin{pmatrix}
    a_{r_{1,\uparrow}, r_{N_\uparrow + 1, \downarrow}} & \ldots & a_{r_{1,\uparrow}, r_{N, \downarrow}}\\
    \vdots & \ddots & \vdots \\
    a_{r_{N_\uparrow}, r_{N_\uparrow + 1, \downarrow}} & \ldots & a_{r_{N_\uparrow}, r_{N, \downarrow}}\\
    \end{pmatrix}.
\end{equation}

It should be noted that we omit all the irrelevant factors in the forms of projective states.

\section{\label{sec:finite_size_effect} FINITE-SIZE EffECT}

\begin{figure}
    \centering
    \includegraphics[width=\linewidth]{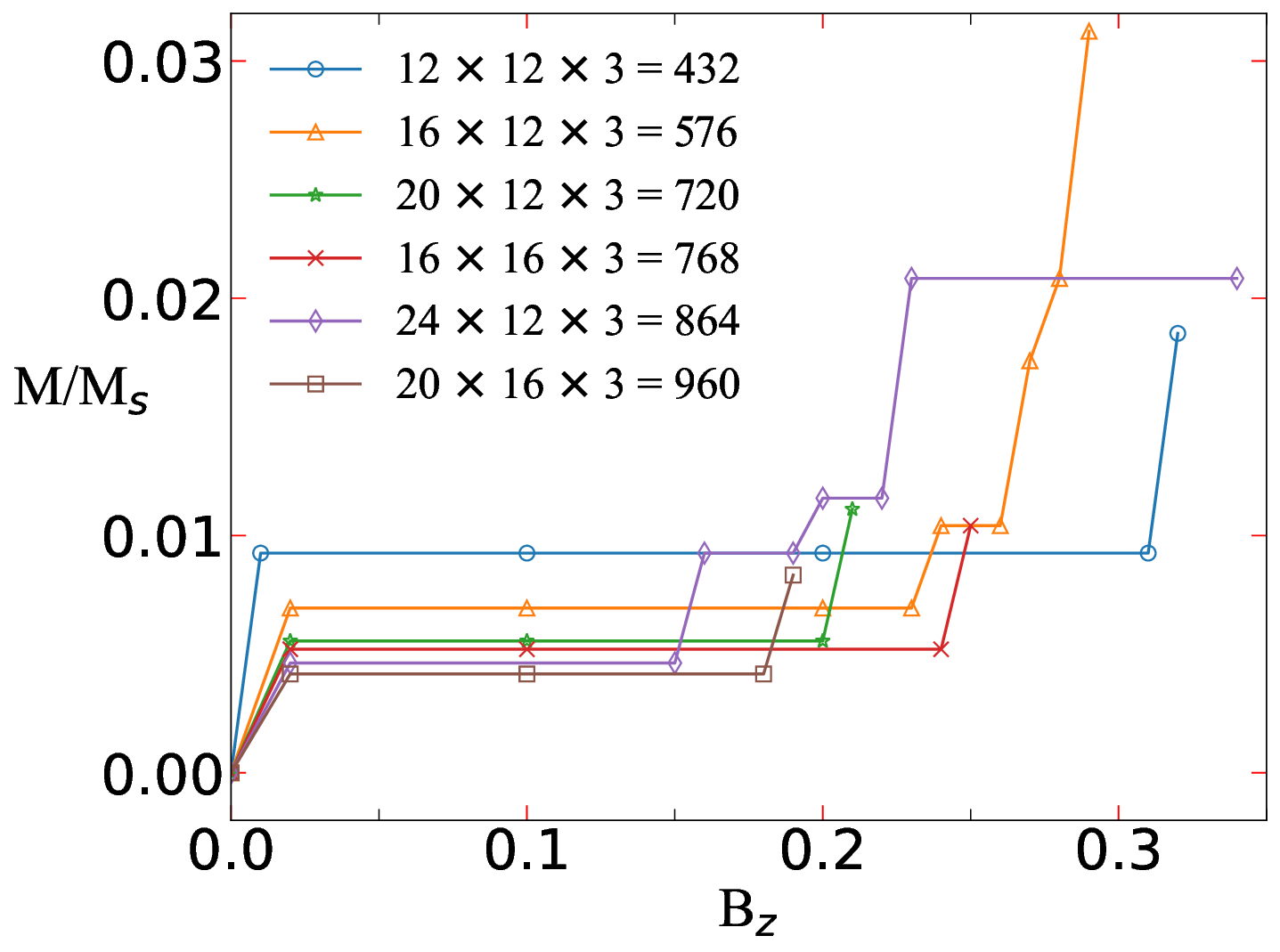}
    \caption{Illustration of the finite-size effect for the average magnetization per site of the DSL when the field $B_z$ is small.}
    \label{fig:m_vs_dsl_size}
\end{figure}

For the VMC technique, the finite-size effect always needs to be considered. In this work, the average magnetization is very sensitive to lattice size. To elaborate this, we study the $M/M_s$ of DSL when the magnetic field $Bz < 0.35$ with different lattice sizes $N$, as shown in Fig.~\ref{fig:m_vs_dsl_size}. It is hard to capture the relatively continuous magnetization for all the sizes we consider. But, this situation for finite-size system is correct because we have to tune the total $S_z$ one by one and the magnetization is always discrete. Besides, as described in main text, a low false magnetization $4/N$ plateau always exists for the gapless DSL. And it must be close to zero with increase of the system size.

\begin{figure}
    \centering
    \includegraphics[width=\linewidth]{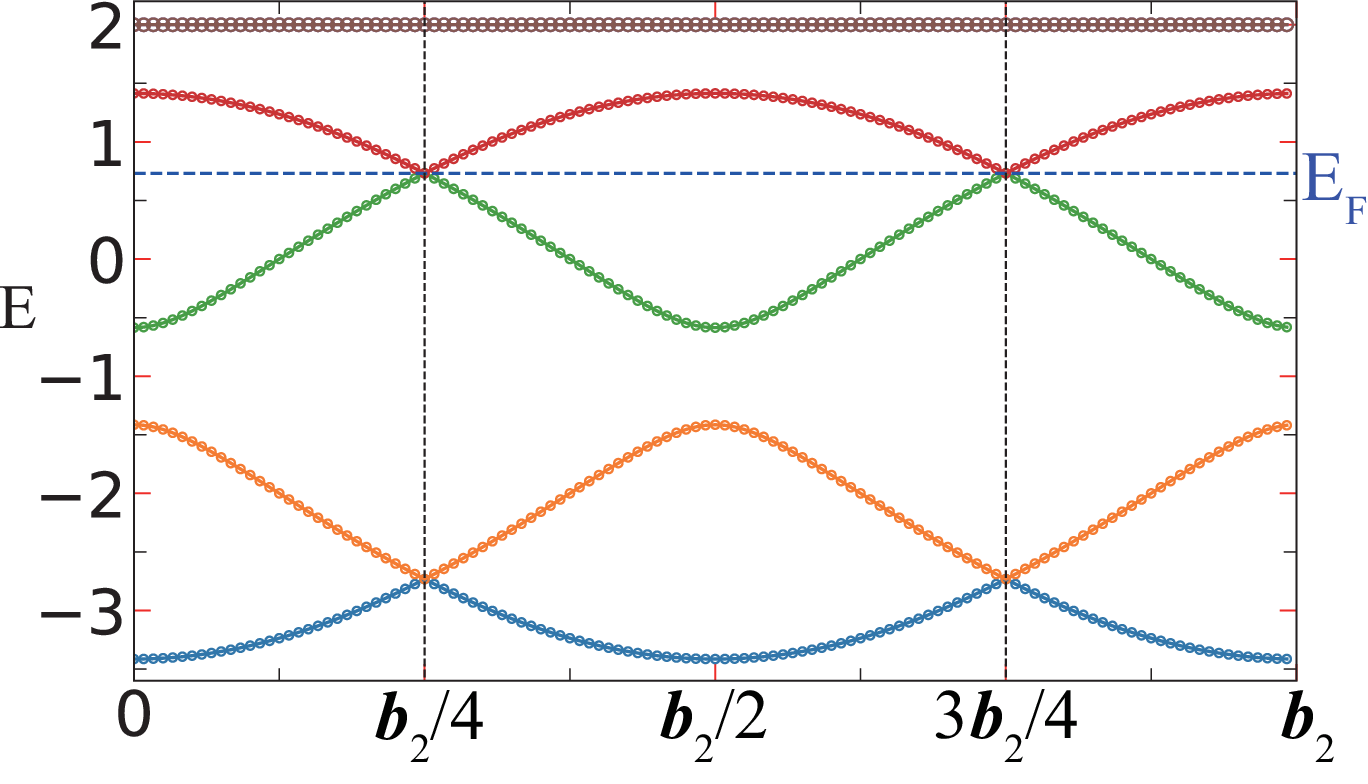}
    \caption{Dispersion of the DSL along with the direction from $\boldsymbol{k} = \boldsymbol{b}_1/2$ to $\boldsymbol{k} = \boldsymbol{b}_1/2 + \boldsymbol{b}_2$ in first BZ. The top flat band is doubly degenerate and the rest bands are non-degenerate. The blue dashed line means the Fermi level. We note we omit the $\boldsymbol{b}_1/2$ in the x-axis label for simplicity.}
    \label{fig:ek_dsl}
\end{figure}

\begin{figure}
    \centering
    \includegraphics[width=\linewidth]{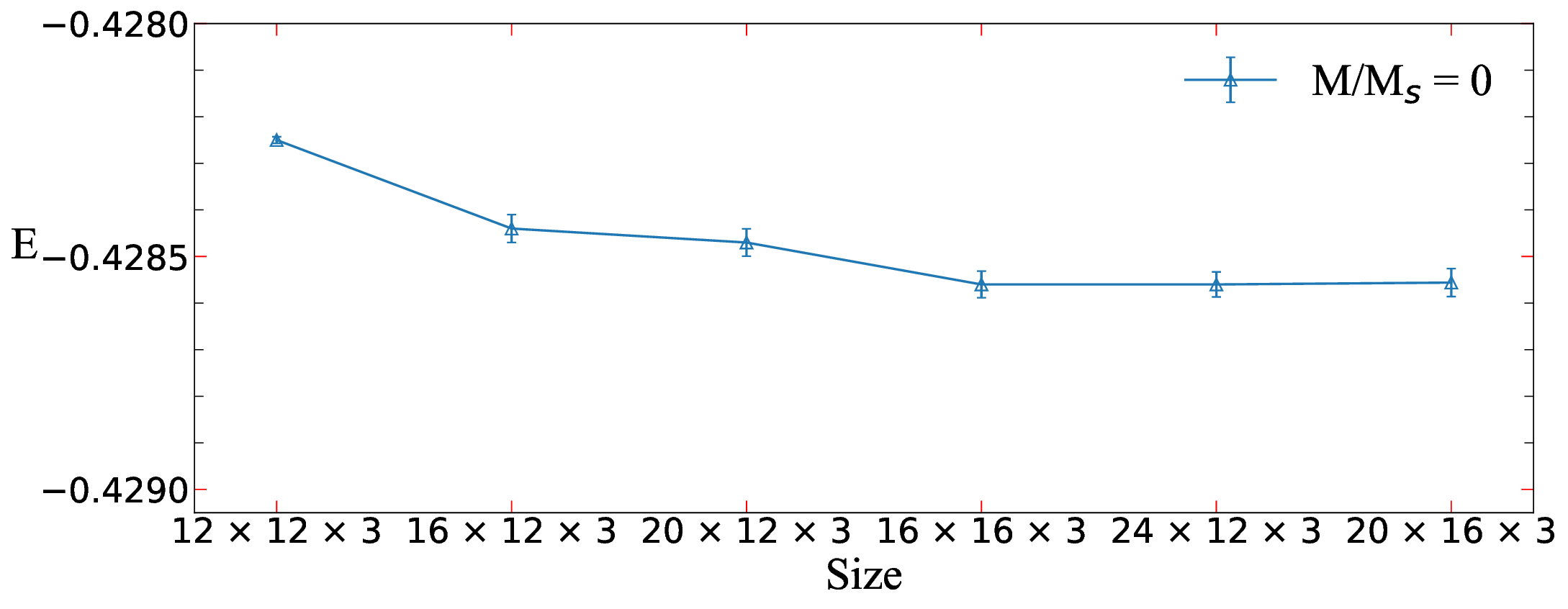}
    \caption{Variational energy per site of the DSL when the average magnetization $M/M_s = 0$ for different lattice sizes.}
    \label{fig:en_vs_size_dsl}
\end{figure}

Now, we explain the reason why we always choose $L_1 = 2n$ and $L_2 = 4n$.  It is because the two Dirac cones in first Brillouin zone (BZ) appear at $\boldsymbol{k}_1 = \boldsymbol{b}_1/2 +  \boldsymbol{b}_2/4$ and $\boldsymbol{k}_2 = \boldsymbol{b}_1/2 +  3\boldsymbol{b}_2/4$ (see Fig.~\ref{fig:ek_dsl}), where $\boldsymbol{b}_{1(2)}$ are the reciprocal bases for the lattice bases $\boldsymbol{a}_1 = (2, 0)$ and $\boldsymbol{a}_2 = (1/2, \sqrt{3}/2)$. In Ref.~\cite{Ran-dsl-2007}, the variational energy per site $E = 0.42866(2)$ for the lattice size $8 \times 8 \times 3$ with the periodic-antiperiodic (P-A) boundary condition (BC). And our calculation indicates $E = 0.42868 \pm 0.00003$ ($E = 0.42724 \pm 0.00003$) for P-A (periodic-periodic (P-P)) BC with $4\times 10 ^5$ samples for the averages, one sample after 240 update with sufficient statistical independent. In fact, the Dirac cones are avoided by the P-A BC. Therefore, it suggest that the variational energy is sensitive to the BC for this kind of size with $L_2 = 4n$. But, to capture the physics of Dirac cones as much as possible, we calculate the variational energy per site with $L_2 = 4n$ and P-P BC for larger sizes, as shown in the Fig.~\ref{fig:en_vs_size_dsl}. Along with the increase of size, the variational energy $E$ is convergent and close to 0.4286. Even for size $16 \times 12 \times 3$, the $E$ is slightly higher than those of the larger sizes, the tendency of magnetization is similar to that of the size $24 \times 12 \times 3$. Therefore, we adopt the size $16 \times 12 \times 3$ for the main results of DSL in main text.

\begin{figure}
    \centering
    \includegraphics[width=\linewidth]{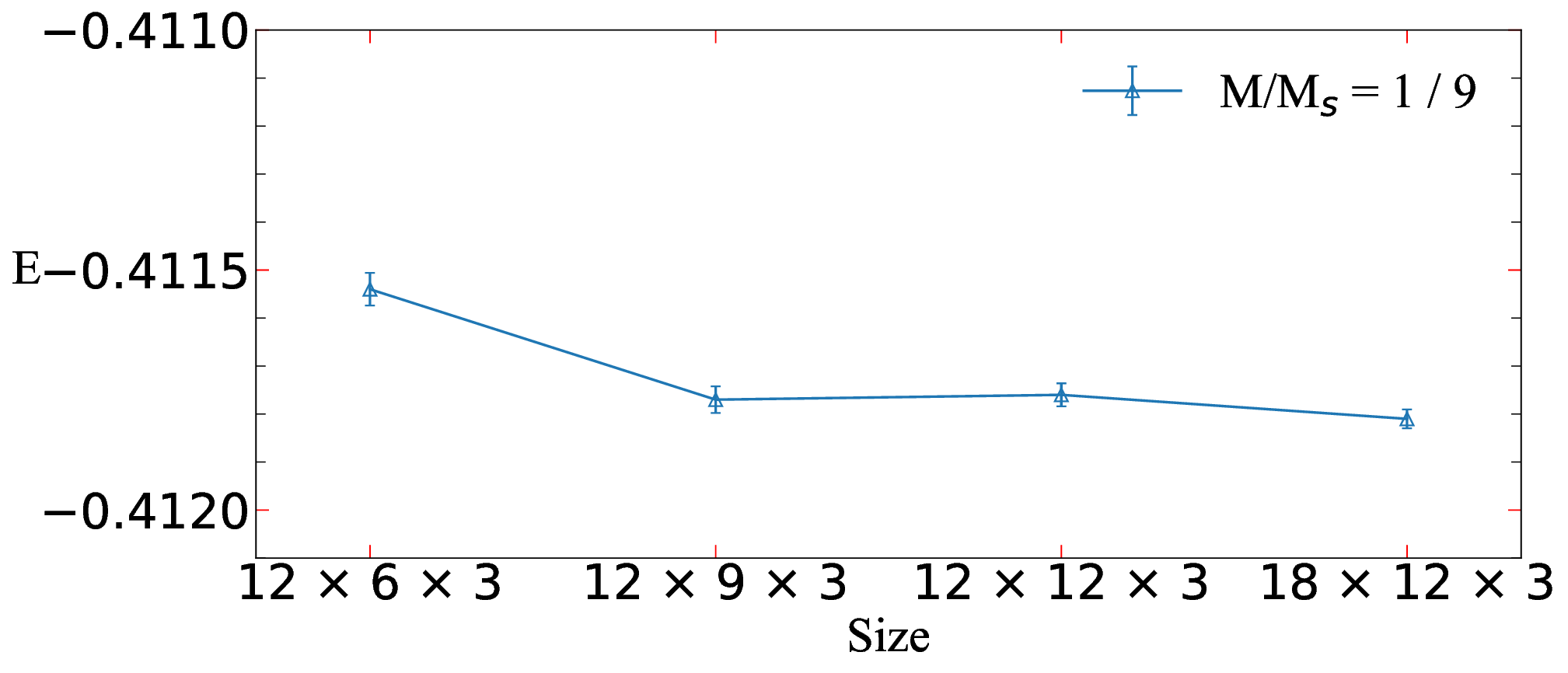}
    \caption{Illustration of the finite-size effect for the variational energy per site of the $\mathbb{Z}_3$ QSL when the average magnetization $M/M_s = 1/9$. And we omit the energy from the field because it is a constant for fixed average magnetization. For the size $18 \times 12 \times 3$, we adopt the optimal parameters from those of size $12 \times 12 \times 3$.}
    \label{fig:en_vs_size_z3}
\end{figure}

For the gapped $\mathbb{Z}_3$ QSL, the variational process becomes highly time-consuming when dealing with very large lattice sizes due to the presence of 15 variational parameters. To examine the finite size effects, we have performed the VMC calculations on four different lattice sizes: $12 \times 6 \times 3$, $12 \times 9 \times 3$, $12 \times 12 \times 3$ and $18 \times 12 \times 3$. The variational energies per site at average magnetization 1/9 plateau are shown in Fig.~\ref{fig:en_vs_size_z3}. We find that the energy differences among the lattices with sizes $12 \times 9 \times 3$, $12 \times 12 \times 3$ and $18 \times 12 \times 3$ are already very small. Therefore, we adopt the size $12 \times 12 \times 3$ for presenting the main results of the $\mathbb{Z}_3$ QSL in the main text.

Additionally, for the $\mathbb{Z}_3$ QSL, the $3\times1$ and $\sqrt{3}\times\sqrt{3}$ extended unit cells are equivalent. As demonstrated in  Table~\ref{tab:en_z3_sqrt3_vs_size}, the variational energies calculated using these two extended unit cells are equal within the margin of error.

\begin{table}[h]
    \centering
    \caption{Variational energy per site omitted the contribution from magnetic field as a function of lattice size for the $\mathbb{Z}_3$ QSL. The second (third) row labeled by $E_{1(2)}$ represents the energy of the $\mathbb{Z}_3$ QSL with the $3 \times 1$ ($\sqrt{3} \times \sqrt{3}$) unit-cell shape. The error bars for all energy values are almost $ \sim 8*10^{-6}$.}
    \begin{tabular}{c|c|c|c}
        \hline \hline
        \ \ \ Size\ \ \  & \ \ \ $12 \times 6 \times 3$\ \ \   & \ \ \ $12 \times 9 \times 3$\ \ \  & \ \ \ $12 \times 12 \times 3$\ \ \  \\
        \hline
        $E_1$ & -0.411520 & -0.411776 & -0.411781 \\
        \hline
        $E_2$ & -0.411506 & -0.411715 & -0.411776\\
        \hline \hline
    \end{tabular}
    \label{tab:en_z3_sqrt3_vs_size}
\end{table}

\section{\label{sec:chern_number} CHERN NUMBER}

A nonzero Chern number is one of the fundamental topological quantities to characterize a topological phase of matter. Here we won't go into details about the concepts of Berry connection, Berry phase and Chern number with formal analytical expression. We just introduce the numerical calculation of the Chern number of the filled bands.

\subsection{\label{subsec:chern_num_mf} Mean-field level}

\begin{figure}
    \centering
    \includegraphics[width=0.95\linewidth]{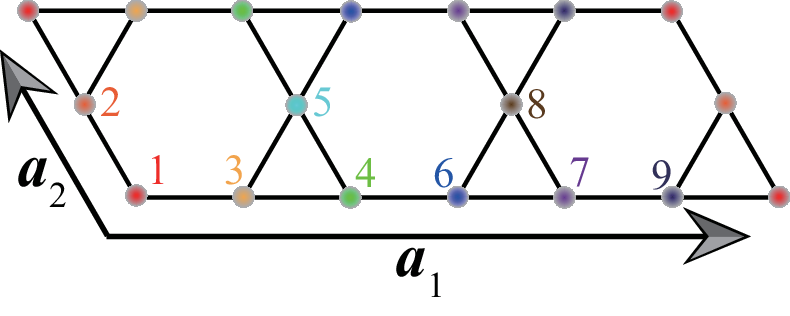}
    \caption{Sublattice indices for the $\mathbb{Z}_3$ QSL. The 9 sites marked in different colors mean 9 different sublattices.}
    \label{fig:chern_z3_ansatz}
\end{figure}

We derive the mean field Hamiltonian using optimized variational parameters from the VMC method, and subsequently transform it into the momentum space to obtain its Bloch form $H(\boldsymbol{k})$. This form is periodic along the directions defined by the reciprocal lattice vectors $\boldsymbol{b}_{1,2}$, satisfying $H(\boldsymbol{k}) = H(\boldsymbol{k} + n_1 \boldsymbol{b}_1 + n_2 \boldsymbol{b}_2)$ for any integers $n_{1,2}$. This Fourier transformation must be handled with care and caution. To be specific, it usually needs another gauge transformation, $c_{\boldsymbol{k}} \longrightarrow c_{\boldsymbol{k}} e^{i\boldsymbol{k} \cdot \boldsymbol{\delta}}$, such as
\begin{equation}
    \begin{aligned}
        &c_{1\boldsymbol{k}} \rightarrow c_{1\boldsymbol{k}},
        c_{2\boldsymbol{k}} \rightarrow c_{2\boldsymbol{k}} e^{i\boldsymbol{k} \cdot \boldsymbol{\delta}_1},
        c_{3\boldsymbol{k}} \rightarrow c_{3\boldsymbol{k}} e^{-i\boldsymbol{k} \cdot 5\boldsymbol{\delta}_2},\\
        &c_{4\boldsymbol{k}} \rightarrow c_{4\boldsymbol{k}} e^{i\boldsymbol{k} \cdot 2\boldsymbol{\delta}_2},
        c_{5\boldsymbol{k}} \rightarrow c_{5\boldsymbol{k}} e^{i\boldsymbol{k} \cdot (\boldsymbol{\delta}_1 + 2\boldsymbol{\delta}_1}),
        c_{6\boldsymbol{k}} \rightarrow c_{6\boldsymbol{k}} e^{i\boldsymbol{k} \cdot 3\boldsymbol{\delta}_2}, \\
        &c_{7\boldsymbol{k}} \rightarrow c_{7\boldsymbol{k}} e^{i\boldsymbol{k} \cdot 4\boldsymbol{\delta}_2},
        c_{8\boldsymbol{k}} \rightarrow c_{8\boldsymbol{k}} e^{i\boldsymbol{k} \cdot (\boldsymbol{\delta}_1 + 4\boldsymbol{\delta}_1)},
        c_{9\boldsymbol{k}} \rightarrow c_{9\boldsymbol{k}} e^{-i\boldsymbol{k} \cdot \boldsymbol{\delta}_2},
    \end{aligned}
    \label{eq:gauge_trans}
\end{equation}
where $\boldsymbol{\delta}_1 = \boldsymbol{a}_2 / 2$ and $\boldsymbol{\delta}_2 = \boldsymbol{a}_1 / 6$. Here, $\boldsymbol{a}_ 1 = (3, 0)$ and $\boldsymbol{a}_ 2 = (-1/2, \sqrt{3}/2)$ are the primitive vector. The sublattice indices are shown in Fig.~\ref{fig:chern_z3_ansatz}

For a lattice with finite size, the Brillouin zone is filled with discrete $\boldsymbol{k}$ points. We define intervals of $\boldsymbol{k}$ points in two directions of reciprocal primitive vectors,
\begin{equation}
    \boldsymbol{u}_i = \frac{l_{i}}{N_{i}}\boldsymbol{b}_{i}, \quad (i = 1,2; \quad N_i / l_i \in \boldsymbol{\mathrm{N}}^* ).
    \label{eq:k_interval}
\end{equation}
In our simulation, we take $l_i = 1$ to guarantee the highest numerical precision. We also note larger intervals are also allowed as long as the result is convergence. We assume that the eigenstate $|n(\boldsymbol{k}) \rangle$ of $H(\boldsymbol{k})$ is also period in Brillouin zone as the same as $H(\boldsymbol{k})$. We can define the U(1) quantity for a certain $\boldsymbol{k}$ as following,
\begin{equation}
    \eta(\boldsymbol{k})_{\boldsymbol{u}_i} \equiv \frac{\langle n(\boldsymbol{k})| n(\boldsymbol{k} + \boldsymbol{u}_i) \rangle}{|\langle n(\boldsymbol{k})| n(\boldsymbol{k} + \boldsymbol{u}_i) \rangle|}.
    \label{eq:e_itheta}
\end{equation}
$\eta(\boldsymbol{k})_{\boldsymbol{u}_i}$ is well defined as long as above denominator is nonzero. Then, we can define another variable about phase in a loop with $\eta(\boldsymbol{k})_{\boldsymbol{u}_i}$,
\begin{equation}
    \begin{aligned}
        &\theta(\boldsymbol{k}) = \frac{1}{i}\ln\left(\eta(\boldsymbol{k})_{\boldsymbol{u}_1} \eta(\boldsymbol{k} + \boldsymbol{u}_1)_{\boldsymbol{u}_2} {\eta(\boldsymbol{k} + \boldsymbol{u}_2)^\dagger_{\boldsymbol{u}_1}} {\eta(\boldsymbol{k})^\dagger_{\boldsymbol{u}_2}}\right),\\
        &-\pi < \theta(\boldsymbol{k}) \leq \pi.
    \end{aligned}
    \label{eq:berry_phase}
\end{equation}
Finally, we define the Chern number which is associated to $n$th band,
\begin{equation}
    C_n \equiv \frac{1}{2\pi} \sum_{\boldsymbol{k} \in \mathrm{BZ}} \theta(\boldsymbol{k}),
    \label{eq:chern_number}
\end{equation}
as shown in main text.

\begin{figure}[htbp]
    \centering
    \includegraphics[width=0.95\linewidth]{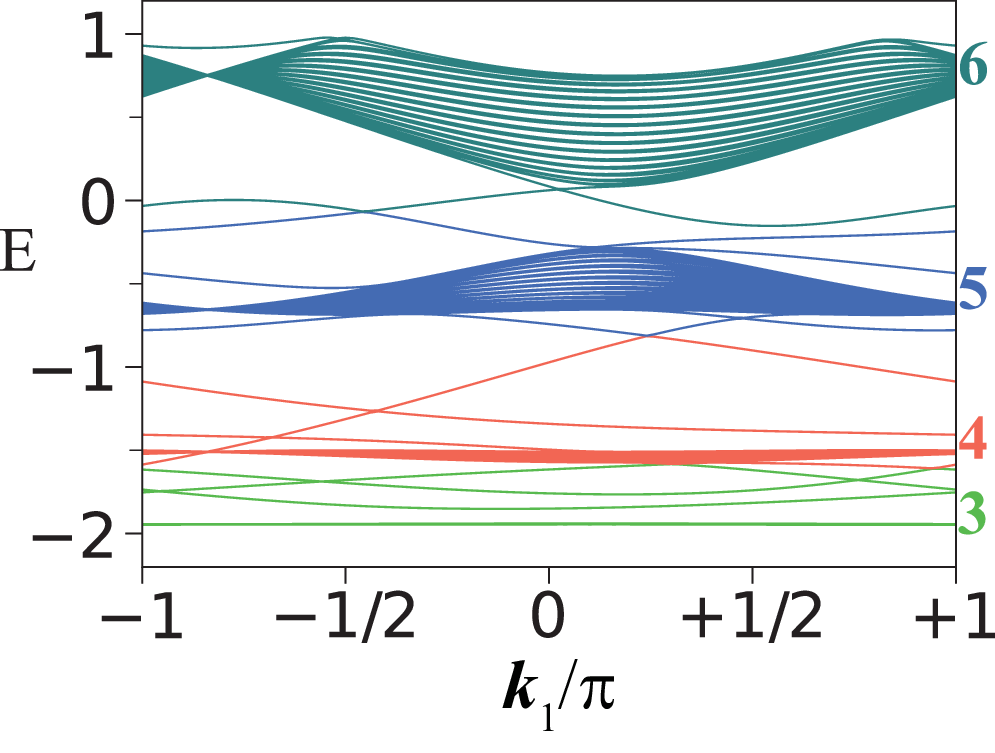}
    \caption{Spinon dispersions (bands 3 to 6) of $\mathbb{Z}_3$ QSL with open boundary condition along the direction of $\boldsymbol{a}_2 = (-1/2, \sqrt{3}/2)$ as shown in Fig.~\ref{fig:chern_z3_ansatz}.}
    \label{fig:edge_state_z3}
\end{figure}

As a consequence of non-trivial Chern number of the exotic $\mathbb{Z}_3$ QSL at the mean-field level, we can observe that the chiral edge states emerge in the gaps for both spin-up and spin-down spinons, as shown in Fig.~\ref{fig:edge_state_z3}.

\subsection{\label{subsec:chern_num_mc} Monte Carlo technique}

First, we need to construct the projective many-body wave function with a spin-dependent twisted boundary condition: $c_{i+L_{j},\uparrow} = c_{i,\uparrow} e^{i\Theta_{j}}$ and $c_{i+L_{j},\downarrow} = c_{i,\downarrow} e^{-i\Theta_{j}}$, where $j \in \{1, 2\}$, $L_{j} = L = 12$ and $\Theta_{j} \in [0, 2\pi]$ are the twisted boundary phases. Here, $j$ indicates the directions of the two primitive lattice vectors, as shown in Fig.~\ref{fig:chern_z3_ansatz}. For numerical calculation, it is hard to capture Berry curvature directly. Therefore, we discretize the space of the phases into a grid consisting of small plaquettes. In our calculations, we adopt $N_{\mathcal{P}} = 10 \times 10 =  100$. The Berry phase of each plaquette $\mathcal{P}$ is given by
\begin{equation}
    \mathrm{BP}_{\mathcal{P}} = \mathrm{Im}\{\ln (\langle \Psi_1 | \Psi_2 \rangle \langle \Psi_2 | \Psi_3 \rangle \langle \Psi_3 | \Psi_4 \rangle \langle \Psi_4 | \Psi_1 \rangle)\},
\end{equation}
where $|\Psi_{1,2,3,4}\rangle$ are the normalized projective many-body function on the four corners of the plaquette $\mathcal{P}$. The overlaps can be calculated by standard Monte Carlo technique, $\langle \Psi_i | \Psi_j \rangle = \sum_x \rho(x) \frac{\langle x | \Psi_j \rangle}{\langle x | \Psi_i \rangle}$, where $\rho(x) = |\langle x | \Psi_i \rangle|^2$ is the sampling weight. Finally, we can calculate the Chern number by summing the Berry phase of each plaquettes, $C = \frac{1}{2 \pi} \sum_{\mathcal{P}} \mathrm{BP}_\mathcal{P}$.

\section{\label{sec:tee} TOPOLOGICAL ENTANGLEMENT ENTROPY}

\begin{figure}
    \centering
    \includegraphics[width=\linewidth]{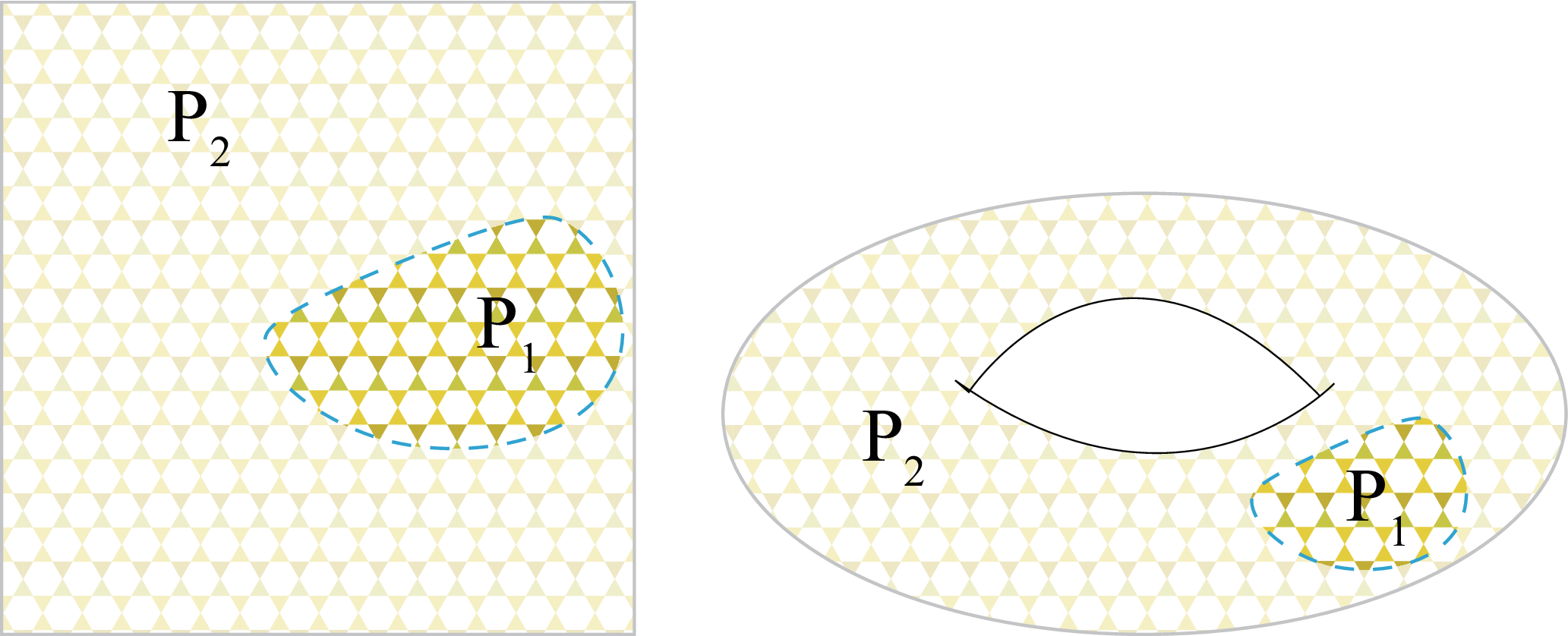}
    \caption{The left panel is a schematic diagram of bipartition on a kagome system and the right one is the corresponding torus. And this bipartition is trivial because the boundaries are contractile.}
    \label{fig:bipartition_and_torus}
\end{figure}

An important quantity to characterize the topological order is the topological entanglement entropy (TEE)~\cite{tee-levin-wen-2006-prl, tee-kitaev-2006-prl}. We divide a system into two parts, as shown in Fig. \ref{fig:bipartition_and_torus}. The von Neumann entanglement entropy of P$_\mathrm{1}$ for an arbitrary system can be represented as follows,
\begin{equation}
    S(L) = \alpha L - \gamma,
    \label{eq:entropy_p1}
\end{equation}
where the coefficient $\alpha$ is not universal and depends on the details of the Hamiltonian, $L$ is the 1-codimensional area of the boundary of P$_\mathrm{1}$ and  $\gamma$ is just the universal TEE. Besides, there is another quantity, topological quantum dimension $D_q$, related to TEE. For instance, for a gapped $\mathbb{Z}_2$ QSL, $\gamma = \ln 2 = \ln D_q$~\cite{Depenbrock-prl-z2}. Namely, we can obtain the quantum dimension of a topological order by its TEE.

Numerical calculation of TEE from the von Neumann entanglement entropy in the VMC method is difficult, so we focus on the Renyi entropy instead of the former one. The Renyi entropy is defined as~\cite{Zhang-tee-2011}
\begin{equation}
    S_n = \frac{1}{1-n} \ln \left[ \mathrm{Tr}\left(\rho_1^n \right) \right],
\end{equation}
where $\rho_1$ is the reduced density matrix by tracing out the subsystem P$_\mathrm{2}$, i.e., $\rho_1 = \mathrm{Tr}_2 |\Psi \rangle \langle \Psi|$, $|\Psi \rangle$ is a normalized wave function of the system. In this paper, we just focus on the Renyi entropy with index $n = 2$, $S_2 = - \ln \left[ \mathrm{Tr}\left(\rho_1^2 \right) \right] = \alpha L - \gamma$. We note that the $\alpha$ in the area term is non-universal and $n$-dependent but the TEE $\gamma$ is universal. In the VMC calculations, we use a swap operator $X$~\cite{renyi-swaps-prl-2010} defined as
\begin{equation}
    X|\alpha_1 \rangle \bigotimes |\alpha_2 \rangle = |\beta_1 \rangle \bigotimes |\beta_2 \rangle,
    \label{eq:swap_operator}
\end{equation}
where $|\alpha_1 \rangle = |a \rangle |b \rangle$, $|\alpha_2 \rangle = |m \rangle |n \rangle$, $|\beta_1 \rangle = |m \rangle |b \rangle$ and $|\beta_2 \rangle = |a \rangle |n \rangle$. Here, $|a \rangle$ and $|m \rangle$ are in P$_\mathrm{1}$, while $|b \rangle$ and $|n \rangle$ are in P$_\mathrm{2}$. We can rewrite $S_2$ in terms of the expectation value of $X$ with respect to the wave function $|\Psi \rangle \bigotimes |\Psi \rangle$, $S_2 = -\ln \langle X \rangle$. We can empirically predict it is a complex number in practical calculation if the $|\Psi \rangle$ is complex. Consequently, we can divide this expectation into two parts, $\langle X \rangle = \langle X_{mod} \rangle \langle X_{phase} \rangle$, which can be individually calculated by Monte Carlo (MC) method. In our VMC calculations, the two key expectation values are given by

\begin{equation}
    \langle X_{mod} \rangle = \sum_{\alpha_1, \alpha_2} \rho_{\alpha_1} \rho_{\alpha_2} \left| f(\alpha_1, \alpha_2) \right|,
\end{equation}
\begin{equation}
    \langle X_{phase} \rangle = \sum_{\alpha_1, \alpha_2} \tilde{\rho}_{\alpha_1, \alpha_2} e^{i\theta(\alpha_1, \alpha_2)},
\end{equation}
\begin{equation}
    \rho_{\alpha_i} = \frac{\left|\langle \alpha_i | \Psi \rangle \right|^2}{\langle \Psi | \Psi \rangle},
        f(\alpha_1, \alpha_2) = \frac{\langle \beta_1 | \Psi \rangle \langle \beta_2 | \Psi \rangle}{\langle \alpha_1 | \Psi \rangle \langle \alpha_2 | \Psi \rangle},
\end{equation}
\begin{equation}
    \tilde{\rho}_{\alpha_1, \alpha_2} = \frac{\left| \langle \alpha_1 | \Psi \rangle \langle \alpha_2 | \Psi \rangle \right|^2 \left| f(\alpha_1, \alpha_2) \right|}{\sum_{\alpha_1, \alpha_2} \left| \langle \alpha_1 | \Psi \rangle \langle \alpha_2 | \Psi \rangle \right|^2 \left| f(\alpha_1, \alpha_2) \right|},
\end{equation}
\begin{equation}
   e^{i\theta(\alpha_1, \alpha_2)} = \frac{f(\alpha_1, \alpha_2)}{\left| f(\alpha_1, \alpha_2) \right|}.
\end{equation}
It should be noted that $\tilde{\rho}_{\alpha_1, \alpha_2}$ is a joint probability distribution related to the detail of plaquette $P_1$. In the main text, to confirm the quantum dimension of the state at 1/9 magnetization plateau, we choose the multiple of primitive cell of kagome lattice, i.e., rhombus, as the contractile plaquette $P_1$ to calculate the entanglement entropy.

\section{\label{sec:gsd} GROUND-STATE DEGENERACY}

The total quantum dimension $D_q$, as mentioned above, is defined as $D_q = \sqrt{\sum_i d_i^2}$ with $d_i$ is the quantum dimension of the $i$th topological excitation. For the Abelian topological phase, the ground-state degeneracy (GSD) satisfies GSD = $D_q^2$. As a canonical example, the toric-code model~\cite{toric_kitaev} supports an Abelian topological phase with four kinds of Abelian (i.e., $d_i = 1$) topological exitations($1, e, m, f$), and GSD = 4.

\begin{figure}
    \centering
    \includegraphics[width=\linewidth]{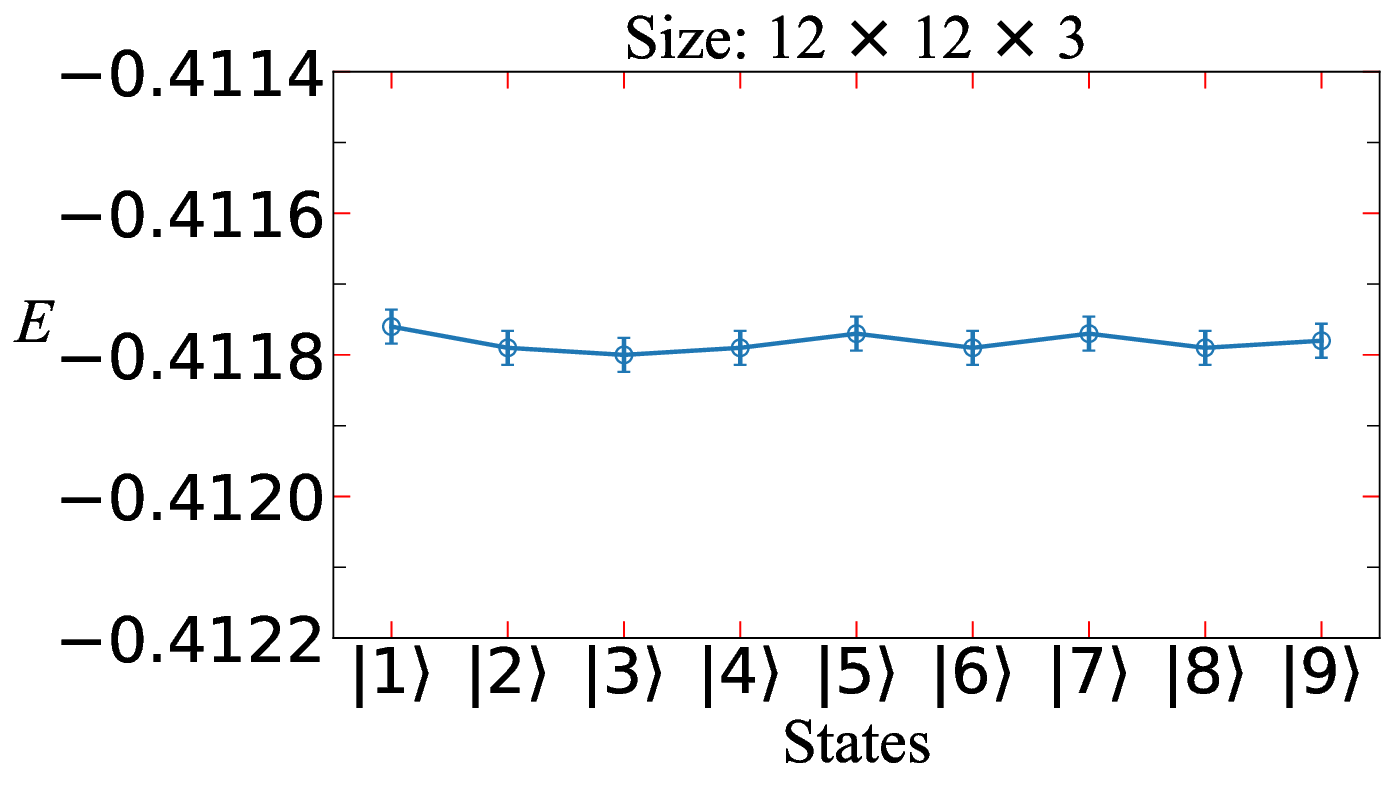}
    \caption{The curve of variational energy (per site, we omit the contribution of external field.) for the $\mathbb{Z}_3$ QSL with different fluxes through the holes of torus at 1/9 magnetization plateau.}
    \label{fig:en_of_9_states}
\end{figure}

The topological order of a gapped QSL can be characterized by its GSD  when one compacts the lattice to a torus: in the thermodynamic limit, there is no energy cost when a $\mathbb{Z}_2$ $\pi$ flux is inserted in any hole of the torus. In the mean field theory, this procedure is equivalent to changing the boundary conditions of $H_\mathrm{mf}$ from period to anti-period. For a 2D system, in general, we can always construct four state $|\phi_{\pm,\pm}\rangle_{\mathrm{mf}}$, where the + (-) subscript means the period (anti-period) boundary condition along the directions of two lattice basis vectors. To extend the analysis to the case of $\mathbb{Z}_3$ QSL, we can attempt to construct 9 states based on $H_\mathrm{mf}$, analogous to the approach for the Z$_2$ QSL, despite we do not know the Abelian or non-Abelian nature of this $\mathbb{Z}_3$ QSL. To be specific, the 9 states are $|\phi_{\alpha, \beta} \rangle_{\mathrm{mf}}$ with $\alpha (\beta) = \{0, 2\pi/3, 4\pi/3 \}$, where $\alpha$ ($\beta$) means the extra phase when the hoppings cross the boundary along the lattice basis $\boldsymbol{a}_1$ ($\boldsymbol{a}_2$) as follows,
\begin{equation}
    \begin{aligned}
        &|1\rangle = |\phi_{0,0}\rangle_{\mathrm{mf}},
        |2\rangle = |\phi_{0,2\pi/3}\rangle_{\mathrm{mf}},
        |3\rangle = |\phi_{0,4\pi/3}\rangle_{\mathrm{mf}}, \\
        &|4\rangle = |\phi_{2\pi/3,0}\rangle_{\mathrm{mf}},
        |5\rangle = |\phi_{2\pi/3, 2\pi/3}\rangle_{\mathrm{mf}},
        |6\rangle = |\phi_{2\pi/3, 4\pi/3}\rangle_{\mathrm{mf}}, \\
        &|7\rangle = |\phi_{4\pi/3,0}\rangle_{\mathrm{mf}},
        |8\rangle = |\phi_{4\pi/3, 2\pi/3}\rangle_{\mathrm{mf}},
        |9\rangle = |\phi_{4\pi/3, 4\pi/3}\rangle_{\mathrm{mf}}.
    \end{aligned}
    \label{eq:9_state_gsd}
\end{equation}
Then, a Gutzwiller projection is required to enforce these 9 ground states to recover physical Hilbert space. As shown in Fig.~\ref{fig:en_of_9_states}, their variational energies of these 9 states should be degenerate within numerical error. Namely, there is no energy cost when a $\mathbb{Z}_3$ flux is inserted in any hole of the torus for this $\mathbb{Z}_3$ QSL at 1/9 magnetization plateau.

We can calculate the 9 by 9 overlap matrix $\mathcal{O}$ with the above 9 states. In detail, the matrix element, $\mathcal{O}_{ij} = \langle i | j \rangle / \sqrt{\langle i | i \rangle \langle j | j \rangle}$, where $i,j = \{1, 2, 3, 4, 5, 6, 7, 8, 9\}$. Obviously, the diagoanl elements $\mathcal{O}_{ii} = 1$ and $\mathcal{O}_{ij} = \mathcal{O}_{ji}^*$. Thus, we calculate the 27 upper-triangular ($i < j$) elements of the overlap matrix with lattice size $12 \times 12 \time 3$ for the $\mathbb{Z}_3$  QSL at 1/9 magnetization plateau as follows,
\begin{align}
&\mathcal{O}_{1,2} = 0.378e^{i2.72}, \mathcal{O}_{1,3} = 0.381e^{i2.59}, \mathcal{O}_{1,4} = 0.384e^{-i2.28}, \nonumber \\
&\mathcal{O}_{1,5} = 0.139e^{i0.61}, \mathcal{O}_{1,6} = 0.378e^{i2.05}, \mathcal{O}_{1,7} = 0.391e^{-i2.28}, \nonumber \\
&\mathcal{O}_{1,8} = 0.384e^{i1.63}, \mathcal{O}_{1,9} = 0.139e^{-i0.50}, \mathcal{O}_{2,3} = 0.382e^{-i0.16}, \nonumber \\
&\mathcal{O}_{2,4} = 0.372e^{i1.29}, \mathcal{O}_{2,5} = 0.373e^{-i2.19}, \mathcal{O}_{2,6} = 0.137e^{-i0.73}, \nonumber \\
&\mathcal{O}_{2,7} = 0.139e^{-i0.46}, \mathcal{O}_{2,8} = 0.387e^{-i1.13}, \mathcal{O}_{2,9} = 0.389e^{i3.02}, \nonumber \\
&\mathcal{O}_{3,4} = 0.143e^{i1.40}, \mathcal{O}_{3,5} = 0.389e^{-i2.02}, \mathcal{O}_{3,6} = 0.391e^{-i0.59}, \nonumber \\
&\mathcal{O}_{3,7} = 0.381e^{-i0.29}, \mathcal{O}_{3,8} = 0.142e^{-i0.99}, \mathcal{O}_{3,9} = 0.379e^{-i3.13}, \nonumber \\
&\mathcal{O}_{4,5} = 0.36e^{i2.85}, \mathcal{O}_{4,6} = 0.375e^{-i1.96}, \mathcal{O}_{4,7} = 0.356e^{-i1.72}, \nonumber \\
&\mathcal{O}_{4,8} = 0.140e^{-i2.37}, \mathcal{O}_{4,9} = 0.360e^{i1.83}, \mathcal{O}_{5,6} = 0.393e^{i1.37},  \nonumber \\
&\mathcal{O}_{5,7} = 0.363e^{i1.62}, \mathcal{O}_{5,8} = 0.377e^{i0.96}, \mathcal{O}_{5,9} = 0.139e^{-i1.10}, \nonumber \\
&\mathcal{O}_{6,7} = 0.138e^{i0.27}, \mathcal{O}_{6,8} = 0.381e^{-i0.37}, \mathcal{O}_{6,9} = 0.373e^{-i2.45}, \nonumber
\end{align}
\begin{equation}
\mathcal{O}_{7,8} = 0.381e^{-i0.64}, \mathcal{O}_{7,9} = 0.362e^{-i2.83}, \mathcal{O}_{8,9} = 0.39e^{-i2.10}.
\label{eq:overlap_z3}
\end{equation}
Finally, we diagonalize this overlap matrix to obtain its eigenvalues. The number of the significantly finite eigenvalues is just GSD.

\section{\label{sec:spin_cor} CORRELATION FUNCTIONS}

\begin{figure}
    \centering
    \includegraphics[width=\linewidth]{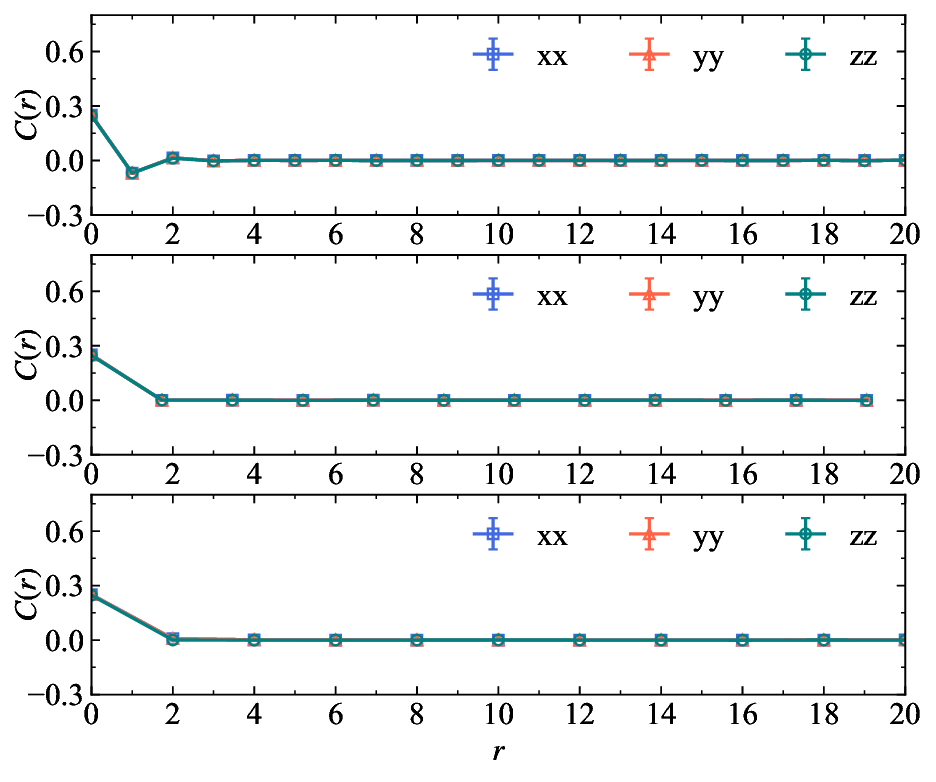}
    \put(-215, 192){(a)}
    \put(-215, 128){(b)}
    \put(-215, 66){(c)}
    \caption{(a), (b) and (c) are the three components of static spin-spin correlation as a function of distance $r$ between two sites along the lines 1, 2 and 3, respectively.}
    \label{fig:spin_cor_xyz}
\end{figure}
\subsection{\label{subsec:}Static correlation functions}
We have elaborated the full static spin correlation of $\mathbb{Z}_3$ topological QSL in the main text. Here, as shown in Fig.~\ref{fig:spin_cor_xyz}, we exhibit its three components for each line as depicted in Fig. 3(b) in the main text. As described in the main text, the spin correlation functions are defined as $C(\boldsymbol{r})_{\alpha\alpha} = \langle \tilde{S}_{\boldsymbol{r}'}^{\alpha} \tilde{S}_{\boldsymbol{r}' + \boldsymbol{r}}^{\alpha}\rangle$, where $\alpha = \{x,y,z\}$ and $\tilde{S}_{\boldsymbol{r}}^\alpha = S_{\boldsymbol{r}}^{\alpha} - \langle S_{\boldsymbol{r}}^{\alpha} \rangle$. We find that the three components of the correlation functions have the same behavior.

\subsection{\label{subsec:sqw} Dynamic structure factors}
Here, we will not delve into the full general technical details of the dynamic spin structure as calculated by the Gutzwiller projected state~\cite{PhysRevB.97.235103, PhysRevB.102.195106, 10.21468/SciPostPhys.14.6.139}, but rather highlight some essential and distinct aspects.

We redefine the dynamic spin structure in the kagome system as follows,
\begin{equation}
\begin{aligned}
    D^{\alpha \beta}(\boldsymbol{q}, \omega)
    &= \sum_{n}\langle\Psi_{G}|\tilde{S}^{z}_{q,\alpha}|\Psi_{n}^{q}\rangle\langle\Psi_{n}^{q}|\tilde{S}^{z}_{q,\beta}|\Psi_{G}\rangle|\\
    &*\delta(\omega-E_{n}^{q}+E_{G}),
\end{aligned}
  \label{eq:sqw_alpha_beta}
\end{equation}
where $\boldsymbol{r}$ and $\boldsymbol{r}^\prime$ are the Bravais vectors, $\alpha$ and $\beta$ are the indexes of sublattice, $E_{G}$ denotes the energy of the ground state $|\Psi_{G}\rangle$, $E_{n}^{q}$ represents the energy of the excited state $|\Psi_{n}^{q}\rangle$ for the momentum $\boldsymbol{q}$, and $\tilde{S}^{z}_{q,\alpha}$ is the Fourier-transformed operator of $\tilde{S}^{z}_{\boldsymbol{r},\alpha}=S^{z}_{\boldsymbol{r},\alpha}-\langle\Psi_{G}|S^{z}_{\boldsymbol{r},\alpha}|\Psi_{G}\rangle$. We note that the reason why we consider the term $\tilde{S}^{z}_{\boldsymbol{r},\alpha}$ is because of the finite magnetization background. For these states with $S^z = 0$, this term goes back to the original form, i.e., $S^{z}_{\boldsymbol{r},\alpha}$. To compare with the observation of inelastic neutron scattering experiment, the total dynamic spin structure is considered by following linear combinations,
\begin{equation}
    D(\boldsymbol{q}, \omega) = \sum_{\alpha, \beta} e^{i \boldsymbol{q} \cdot (\boldsymbol{\delta}_\alpha - \boldsymbol{\delta}_\beta)} D_s^{\alpha \beta}(\boldsymbol{q}, \omega).
    \label{eq:sqw}
\end{equation}



\bibliography{sm_kagome_field}